\documentclass[11pt]{article}

\usepackage[letterpaper,margin=1in]{geometry}

\usepackage{microtype}
\usepackage{amsmath}
\usepackage{amssymb}
\usepackage{amsthm}

\usepackage[authoryear,round]{natbib}

\usepackage{graphicx}
\usepackage{xcolor}
\usepackage{booktabs}
\usepackage{subfig}
\usepackage{enumitem}
\usepackage{mathtools}
\usepackage{bbm}
\usepackage{bm}
\usepackage{upgreek}
\usepackage[title]{appendix}
\usepackage[colorlinks,citecolor=blue,urlcolor=blue,linkcolor=blue]{hyperref}

\newcommand{\subfigure}[2][]{\subfloat[#1]{#2}}

\theoremstyle{plain}
\newtheorem{theorem}{Theorem}[section]

\newtheorem{corollary}{Corollary}[section]

\theoremstyle{definition}
\newtheorem{definition}{Definition}[section]

\newtheorem{assumption}{Assumption}

\theoremstyle{remark}
\newtheorem{remark}{Remark}

\newcommand{\newexample}[1]{%
  \theoremstyle{definition}%
  \newtheorem{example#1}{Example}%
  \expandafter\renewcommand\csname theexample#1\endcsname{#1\Alph{example#1}}%
}
\newexample{1}
\newexample{2}
\newexample{3}

\newcommand{\E}{\mathbb{E}}
\newcommand{\Pb}{\mathbb{P}}

\newcommand{\var}{\text{var}}

\newcommand{\Pn}{\mathbb{P}_n}

\newcommand{\R}{\mathbb{R}}

\newcommand{\dist}{\textsf{dist}}
\newcommand{\sol}{\mathsf{s}^*}

\begin{document}

\title{Semiparametric Inference for Counterfactual Regression under Intervention-Driven Shift}

\author{Kwangho Kim\\
Department of Statistics, Korea University\\
\texttt{kwanghk@korea.ac.kr}}

\date{}

\maketitle

\begin{abstract}
We study counterfactual regression, which maps features to outcomes under
hypothetical scenarios that differ from those observed in the data. This
problem is central to decision-making under distribution shift, where
treatment patterns may change at deployment. We develop a semiparametric
framework for counterfactual regression along a prespecified
incremental-intervention path. The target is a finite-dimensional
constrained projection of counterfactual risk, estimated using
cross-fitted influence-function representations of the program
components. For smooth programs with fixed constraints and
finite-dimensional programs with estimated linear constraints, we
establish consistency and local stability of the optimizer under class-specific conditions, and derive pointwise and uniform first-order expansions.
These results yield asymptotically valid inference, including
simultaneous confidence bands for the counterfactual regression path.
Simulations and an application to SMS reminders illustrate the
finite-sample performance and practical applicability of the proposed
approach.

\medskip
\noindent\textit{MSC2020 subject classifications:} Primary 62G05.

\end{abstract}


\section{Introduction}
\label{sec:intro}

A counterfactual describes what would have happened under a hypothetical intervention. Formalized as potential outcomes, they have long been the standard language of causal inference; more recently, they have also proved useful for prediction under distribution shift, where deployment conditions differ substantially from those seen during training. This is commonly referred to as \textit{counterfactual prediction}. In counterfactual prediction, the goal is to map input features to outcomes under hypothetical scenarios that deviate from observed data (e.g., what if treatment access were limited to 10\% of individuals compared to current practice?). This is central in business and clinical decision-making, where one must anticipate and adapt to abrupt shifts in treatment patterns \citep[e.g.,][]{dickerman2020counterfactual, van2020prediction, dickerman2022predicting}. Moreover, certain interventions in the training data may be undesirable or incompatible with the target deployment setting, requiring their removal via prediction under a hypothetical intervention that completely excludes them \citep[e.g.,][]{sperrin2018using}. This is especially relevant in clinical research when predicting risk as to treatments initiated after baseline \citep{schulam2017reliable, van2020prediction, lin2021scoping}. 

More generally, counterfactual prediction may be used when transferring a model from a source population to a target population that differs only in post-baseline treatment patterns. Therefore, counterfactual prediction is closely related to domain adaptation and out-of-distribution generalization, especially when distributional shifts result from specific interventions. Ideally, one would retrain the model on data from the new setting, but when this is costly or infeasible, an alternative is to adapt the existing model to approximate outcomes under a hypothetical intervention consistent with treatment patterns in the target setting.

Counterfactual prediction is harder than standard prediction because the training signal is intrinsically unobserved, requiring identification through causal structure rather than direct supervision. While modern prediction methods have fueled progress in causal inference, using causal tools to tackle predictive problems under intervention-driven shift remains less developed. Although related to conditional average treatment effect (CATE)
estimation, which contrasts counterfactual regressions, counterfactual
prediction instead targets the regime-specific regression functions
themselves \citep{kennedy2020optimal}. In multi-valued treatment problems, this also makes direct estimation of each counterfactual regression function preferable to enumerating all pairwise contrasts.

Some recent progress has been made for counterfactual prediction beyond its role within the CATE estimation framework. To address covariate distributional shifts induced by varying treatments, several studies have adopted direct modeling, or plug-in, approaches
grounded in the parametric g-formula
\citep[e.g.,][]{li2016predictive,schulam2017reliable,
nguyen2020counterfactual,lin2021scoping,dickerman2022predicting}. \citet{coston2020RuntimeConfounding} introduced a nonparametric method for counterfactual prediction and associated risk, addressing runtime confounding. Other approaches utilize representation learning techniques to construct a common representation space that balances the source and target domains \citep[e.g.,][]{johansson2016learning, shalit2017estimating, hassanpour2019counterfactual, hassanpour2019learning}. \citet{mcclean2024nonparametric} proposed nonparametric methods to estimate counterfactual outcomes under incremental interventions. Yet, no general framework exists for efficient counterfactual prediction under broad risk criteria and flexible constraints.

Our work aims to fill this gap. We develop a semiparametric framework
for counterfactual regression under prespecified intervention-driven
shifts, using incremental interventions
\citep{kennedy2019nonparametric} and allowing flexible risk criteria
and fixed or estimated linear constraints.
\citet{mcclean2024nonparametric} develop nonparametric estimators for
related conditional incremental-effect targets in an unconstrained
setting, whereas our framework incorporates constraints on the
counterfactual regression rule and provides uniform inference over the
intervention path. The estimand is a finite-dimensional, possibly constrained
counterfactual risk minimizer, defined as the optimizer of a stochastic
program. Under squared-error loss, it corresponds to a penalized and
constrained projection of
$x\mapsto\E\{Y^{Q(\delta)}\mid X=x\}$. We estimate it through an approximating
program whose unknown components are constructed using efficient
influence functions and cross-fitting. For smooth programs with fixed
constraints and finite-dimensional programs with estimated linear
constraints, we establish consistency and local stability under
class-specific conditions and derive pointwise and uniform first-order
expansions. Simulations demonstrate robustness to nuisance estimation,
and an application to short message service (SMS) reminders
illustrates counterfactual risk profiling under shifted reminder intensity.

\section{Framework}

\subsection{Setup}
Suppose that we have access to an i.i.d. sample $(Z_{1}, ... , Z_{n})$ of $n$ tuples $Z=(Y,A,X) \sim \Pb$,
where $Y \in \mathcal{Y}$ represents the outcome, $A \in \{0,1\}$ denotes an intervention, and $X \in \mathcal{X} \subseteq \R^p$ comprises observed covariates. Here, we assume $A$ is binary for simplicity, but in principle it can be multi-valued. We consider $\mathcal{Y}$ as a subset of Euclidean space, though it may also be discrete. We let $Y^a$ denote the counterfactual outcome that would be observed under treatment assignment $A = a$, $a \in \mathcal{A}$. We refer to prediction of the observable outcome $Y$ from a set of
predictors as \emph{factual prediction}, in contrast to counterfactual
prediction.

Throughout, we assume the standard causal identification assumptions of consistency and no unmeasured confounding \citep[][Chapter 12]{imbens2015causal}:
$
Y=Y^A
$ and 
$
Y^a\perp A\mid X
$, $\forall a \in \mathcal{A}$. We employ incremental interventions to generate our target counterfactual scenarios with greater flexibility. \emph{Incremental interventions} are a specialized type of stochastic intervention that shifts the exposure propensity of each unit by a predetermined amount \citep{kennedy2019nonparametric}. Specifically, let $\pi(X)=\Pb(A=1 \mid X)$ denote the observational propensity score. Then the incremental intervention replaces $\pi$ with the distribution defined by
\begin{align*}
    q(X; \delta, \pi) = \frac{\delta \pi(X)}{\delta \pi(X) + 1 - \pi(X)} \quad \text{for} \quad 0 < \delta < \infty,
\end{align*}
where $\delta$ is a user-specified increment parameter, indicating  how the intervention changes the odds of receiving treatment; if $\delta > 1$ ($< 1)$, then the intervention increases (decreases) the odds. This corresponds to a counterfactual scenario in which individuals receive a hypothetical treatment sampled from $\text{Bernoulli} \{ q(X; \delta, \pi) \}$. We use $\{Q(\delta):\delta>0\}$ as a prespecified sensitivity path through stochastic treatment regimes, allowing smooth changes in treatment propensity around the observed treatment mechanism. For finite $\delta$, these interventions preserve the support of the observed treatment mechanism and avoid the strict positivity conditions of deterministic interventions. This intervention class has been studied and extended in a range of causal settings, including longitudinal studies with dropout \citep{kim2021incremental}, conditional incremental effects \citep{mcclean2024nonparametric}, and continuous exposures \citep{schindl2024incremental}. When a target marginal treatment rate $p_\star$ has a direct operational interpretation, $\delta$ may be calibrated, when feasible, by solving
\[
\E\{q(X;\delta,\pi)\}=p_\star.
\]
Thus, $\delta$ may be interpreted either as an odds-shift sensitivity parameter or, after calibration, through the corresponding marginal intervention rate. We also allow the deterministic limits $\delta\in\{0,\infty\}$, which recover $A=0$ and $A=1$ under the additional positivity condition: $\min\{\pi(X),1-\pi(X)\}\geq\pi_{\min}>0
\quad\text{a.s.}$.

Let $Q(\cdot\mid X;\delta)$ be the incremental-intervention distribution induced by
$q(X;\delta,\pi)$, and write $Y^{Q(\delta)}$ for the corresponding counterfactual outcome.
Under consistency and no unmeasured confounding, $\E\{Y^{Q(\delta)}\}$ admits the identification formula:
\begin{align}
\E\{Y^{Q(\delta)}\}
&=
\E\!\left[
\frac{\delta \pi(X)\mu_1(X)+\{1-\pi(X)\}\mu_0(X)}
{\delta \pi(X)+\{1-\pi(X)\}}
\right]
=
\E\!\left[
\frac{Y(\delta A+1-A)}{\delta \pi(X)+\{1-\pi(X)\}}
\right],
\label{eqn:mean-inc-outcome-identification}
\end{align}
where $\mu_a(X)=\E(Y\mid X,A=a)$.
Our goal is not the marginal mean but counterfactual regression, predicting $Y^{Q(\delta)}$ from $X$; for $\delta$ far from $1$, factual regression can be inaccurate because the intervention reweights the treated and control regressions.

Incremental interventions are related to treatment choice problem \citep[e.g.,][]{kitagawa2018should, kitagawa2023treatment} in that both allow treatment intensity to vary at the population level. The distinction is that treatment choice problems typically optimize over the treatment fraction or assignment rule, whereas incremental interventions provide a stochastic regime for counterfactual prediction and inference that perturbs the observed treatment mechanism and helps avoid the strong positivity requirements of deterministic regimes; extending our framework to optimize over $\delta$ or over a class of stochastic rules is an interesting direction for future work.

\textbf{Notation.}
For any fixed vector $v$ and matrix $M$, we let $\Vert v \Vert_2$ and $\Vert M \Vert_F$ denote the Euclidean norm and the Frobenius norm, respectively. $\Vert \cdot \Vert_2$ is understood as the spectral norm when it is used with a matrix. Let $\Pn$ denote the empirical measure over $(Z_1,...,Z_n)$. In addition, we use $\Vert f \Vert_{2,\Pb}$ to denote the $L_2(\Pb)$ norm of $f$ defined by $\Vert f \Vert_{2,\Pb} = \left[\int f(z)^2 d\Pb(z)\right]^{1/2}$. Lastly, we let $\sol(\textsf{P})$ denote the set of optimal solutions of an optimization program $\textsf{P}$, and define $\dist(x, \mathcal{S}) = \inf\left\{ \Vert x - y\Vert_2: y \in \mathcal{S} \right\}$ as the distance from a point $x$ to a set $S$.

\subsection{Motivating Illustration} \label{subsec:motivating-illustration}

Consider a simple data-generating process: $X \sim \text{Uniform}[0, 10]$,
$A \sim \text{Bernoulli}(\text{expit}(2.8 - 0.3X))$, and $Y \sim N(1 + 0.75X + 0.5X^2 - 10A, X)$, where $\text{expit}(\cdot)$ denotes the inverse logit function.
Under this model, approximately 75.4\% of individuals initiate treatment. However, treatment becomes less likely as $X$ increases, so a policy that sharply reduces treatment intensity induces a pronounced shift in $\Pb$.

We consider counterfactual prediction under an incremental intervention with $\delta\in\{0.1,0.01\}$, which provides sensitivity scenarios corresponding to substantial reductions in access (e.g., ``for a patient with baseline predictors $X=x$, what would be the 5 year risk of death if access to heart transplant were substantially more limited than in the training population?"; \citealp{dickerman2020counterfactual}).
We generate independent training and test samples of size $n=1000$, estimate the propensity score by logistic regression, and in the test sample draw $A\sim\text{Bernoulli}\{q(X;\delta,\pi)\}$.
We compare (i) a factual regression fit by ordinary least squares (with polynomial terms) and (ii) our counterfactual regression estimator using $L_2$ loss without constraints, under correct model specification.

\begin{figure}[t!]
\centering
\subfigure[]{\includegraphics[width=0.34\linewidth]{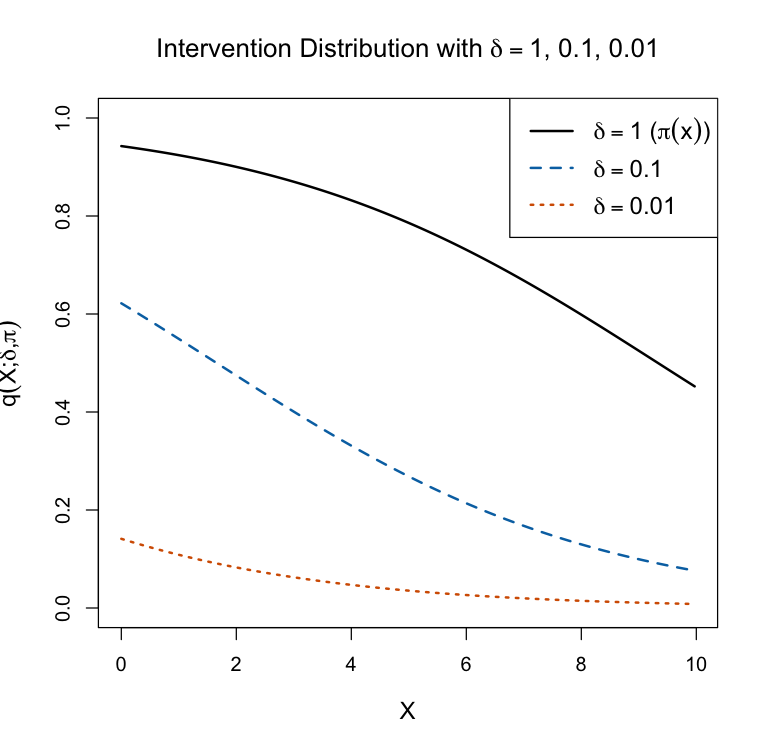}} 
\hfill%
\subfigure[]{\includegraphics[width=0.32\linewidth]{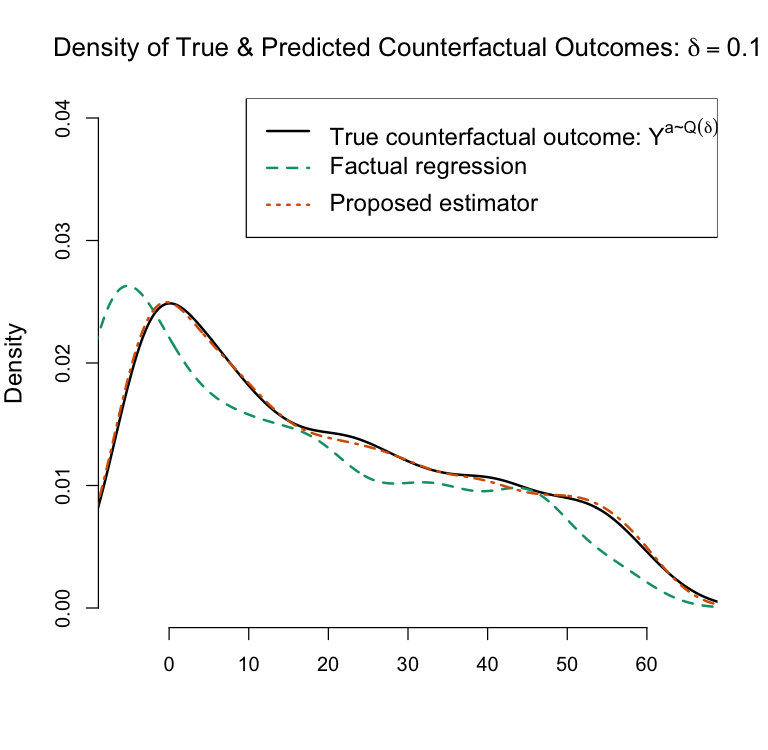}}
\hfill%
\subfigure[]{\includegraphics[width=0.32\linewidth]{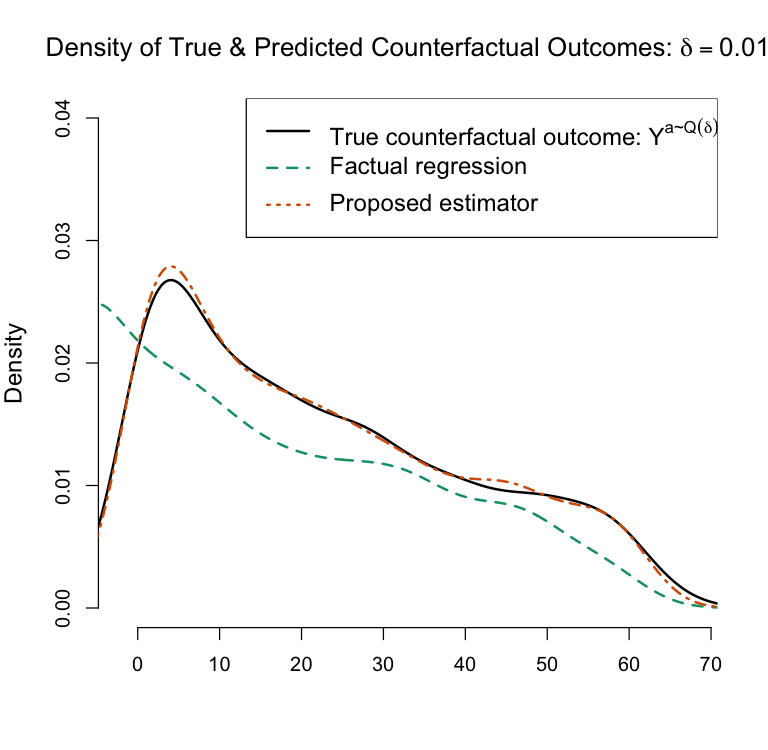}}
\hfill%
\caption{(a) Incremental intervention probability $q(X;\delta,\pi)$ as a
function of $X$ for varying values of $\delta$; (b) and (c) display the densities of the true counterfactual outcomes and the predicted outcomes from factual and proposed counterfactual regression methods for $\delta=0.1$ and $\delta=0.01$, respectively. The factual regression exhibits a notable distributional discrepancy from the true counterfactual outcomes, resulting in substantial estimation errors.
}
\label{fig:illustrating-example}
\end{figure}

Figure~\ref{fig:illustrating-example} shows that as $\delta$ decreases, the induced treatment distribution differs sharply from the observational one, and factual regression no longer matches the counterfactual outcome distribution.
In contrast, the proposed method tracks the counterfactual distribution
well across $\delta$ in this correctly specified illustration. 

\subsection{Estimand} \label{subsec:estimand}

Learning tasks are generally characterized through risk functions. Across domains, a wide variety of nontraditional risk functions are often employed to design and evaluate learning algorithms \citep{wang2020comprehensive}. 
Given a loss function $L$, we define the risk of a regression model $f:\mathcal{X} \rightarrow \R$ as $\mathcal{R}\left( f(X), Y^{Q(\delta)} \right)
=\E\left\{L\left(Y^{Q(\delta)},f(X)\right)\right\}$. When the $L_2$ loss (or squared error loss) is used without any model constraints, the function $f$ that minimizes $\mathcal{R}(f(X), Y^{Q(\delta)})$ is given by the conditional mean $\mathbbm{E} \{ Y^{Q(\delta)} \mid X \}$ \citep{mcclean2024nonparametric}. We define the regression model for $Y^{Q(\delta)}$ as a parametric family $\{f(x;\beta) : \beta \in \mathbb{R}^k\}$; consequently, the target parameter is given by the optimal solution to the following stochastic program:
\begin{align} \label{eqn:target-program}
\underset{\beta \in \R^k}{\text{minimize }} \,\,  \mathcal{R}\left( f(X;\beta), Y^{Q(\delta)} \right) + \lambda \rho(\beta) \quad \text{subject to} \quad \beta \in \mathcal{C}.  
\end{align}
Here, the constraint set is $\mathcal{C} = \{\beta : g_j(\beta) \leq 0, j=1,\ldots,r\}$ for $g_j:\R^k \rightarrow \R$, $\rho(\cdot)$ is a pre-specified penalty function, and $\lambda \geq 0$ is the associated tuning parameter.
The optimizer $\beta^*(\delta)$ defines the best-in-class
counterfactual prediction rule under the specified loss, penalty, and
constraints. Under squared-error loss, the unconstrained functional
target is $m_\delta(x)=\E\{Y^{Q(\delta)}\mid X=x\}$, so $\beta^*(\delta)$ is a penalized and constrained
finite-dimensional $L_2$ projection of $m_\delta$. Other losses yield
the corresponding conditional loss-optimal target. Thus,
$x\mapsto f\{x;\beta^*(\delta)\}$ provides a covariate-specific summary
of outcomes under the intervention regime, enabling risk stratification
and prediction subject to structural or operational constraints---information
that is unavailable from the marginal counterfactual mean
$\E\{Y^{Q(\delta)}\}$ alone.

More concretely, for a fixed $\delta$, the score
$x\mapsto f\{x;\beta^*(\delta)\}$ ranks covariate profiles according to
their fitted counterfactual target under $Q(\delta)$ and can therefore
inform risk stratification when larger values represent greater risk.
When interest instead concerns changes across intervention regimes, one
may compare regime-specific conditional means or their corresponding
finite-dimensional projections; see
Remark~\ref{rmk:conditional-effects}. Thus, the framework distinguishes
risk profiling within a regime from covariate-specific comparisons
across regimes.

Because we impose no assumptions on the true relationship between $Y^{Q(\delta)}$ and $X$, our results are formally nonparametric, regardless of the parametric class $\{f(x;\beta): \beta \in \mathbb{R}^k\}$ employed for estimation. The above projection approach, where a parametric model is not assumed to be correct but is instead used solely to define approximations, has been widely employed in statistics and causal inference \citep[see, for example,][Section 3.1.1]{kennedy2021semiparametric}. Here, we treat $\lambda$ as a user-chosen regularization level that defines the projection target, and hence the estimand, in \eqref{eqn:target-program}. In practice, $\lambda$ may be selected by cross-validation using an estimated counterfactual risk (e.g., a cross-fitted criterion based on efficient influence functions). Our subsequent asymptotic theory is stated for fixed $\lambda$; extending it to data-adaptive $\lambda$ requires standard stability conditions on the selection rule and the empirical criterion. Our framework also allows prediction to depend only on a subset $V\subseteq X$, accommodating settings such as runtime confounding where some covariates are available for nuisance learning but unavailable at prediction time \citep{coston2020RuntimeConfounding, kim2022counterfactual}.

\textbf{Models.}
We present a few illustrative examples of model classes for $f(x; \beta)$.
An archetypal example is a class of generalized linear models, where the counterfactual outcome has a distribution that is in the exponential family. More generally, one may consider a functional aggregation or linear ensemble in the following form:
\begin{align} \label{eqn:function-aggregation}
    f(X;\beta) = \sum_{j=1}^k \beta_j b_j(X),
\end{align}
with a set of known basis functions $\{b_j\}$. Examples of such basis functions include orthogonal polynomials, wavelets, and splines. Aggregated predictors of this form have been extensively studied and applied across a wide range of disciplines \citep[e.g.,][]{tsybakov2003optimal, nemirovski2009robust, polley2010super}. 

\textbf{Constraints.}
We often seek to incorporate constraints into regression models so that the resulting prediction rule respects structural knowledge, scientific principles, or operational requirements. In our setting, these constraints restrict the admissible counterfactual regression functions $x\mapsto f(x;\beta)$ under the prespecified regime $Q(\delta)$, rather than defining a treatment allocation rule. A few examples that may be included in the constraint set $\mathcal{C}$ are presented below. Some constraints are fixed by design, whereas others involve population quantities and are therefore estimated from the data.

\begin{example1}[Constrained counterfactual regression]\label{example:constrained-regression}
In counterfactual regression, shape or structural constraints may be imposed when scientific knowledge suggests that the predicted counterfactual outcome should vary in a particular direction with a covariate.
For instance, if $Y^{Q(\delta)}$ is an adverse event risk, one may require the predicted risk to be nondecreasing in disease severity, exposure duration, or waiting time. With a linear aggregation $f(X;\beta)=\sum_{j=1}^k\beta_j b_j(X)$, many such restrictions can be encoded as linear inequality constraints $C\beta\le d$ (with $C\in\R^{r\times k}$ and $d\in\R^r$). Similar constrained regression formulations have been used to enforce shape constraints \citep{james2013penalized,gaines2018algorithms}, structural consistency \citep{liew1976inequality}, physical calibration requirements \citep{schwendinger2024holistic}, and compositional restrictions \citep{lu2019generalized}.
The resulting estimator is a constrained counterfactual prediction rule
under the specified risk, rather than a treatment allocation rule.
\end{example1}

\begin{example1}[Fair counterfactual scoring]\label{example:fairness}
In some applications, the counterfactual prediction score itself may be required to satisfy fairness or parity constraints across sensitive groups. Let $F=F(X)\in\{0,1\}$ be a binary sensitive feature, and let $f(X;\beta)$ denote the predicted counterfactual outcome under the prespecified regime $Q(\delta)$. A broad class of group fairness constraints can be expressed through moment restrictions on the prediction score \citep[e.g.,][]{mishler2021fade, kim2023fair}. Specifically, for a user specified fairness function $u_f$, we may require
\begin{align}
\left| \E\left\{u_f(Z) f(X;\beta)\right\} \right| \leq \epsilon,
\label{eqn:fairness-function}
\end{align}
where $\epsilon\geq0$ is an acceptable fairness threshold. This constraint restricts the counterfactual regression function $x\mapsto f(x;\beta)$, rather than defining a treatment allocation rule. For example, statistical parity of the counterfactual score can be obtained by setting
\begin{align}
u_f(Z)
=
\frac{1-F}{\E(1-F)}
-
\frac{F}{\E(F)}.
\label{eqn:independence}
\end{align}
Then \eqref{eqn:fairness-function} becomes
\begin{align}
\left|
\E\{f(X;\beta)\mid F=0\}
-
\E\{f(X;\beta)\mid F=1\}
\right|
\leq
\epsilon .
\label{eqn:statistical-parity-score}
\end{align}
Similarly, conditional statistical parity given a legitimate factor $S=S(X)$ can be encoded by setting
\begin{align}
u_f(Z)
=
\frac{(1-F)\mathbbm{1}\{S=s\}}
{\E\left[(1-F)\mathbbm{1}\{S=s\}\right]}
-
\frac{F\mathbbm{1}\{S=s\}}
{\E\left[F\mathbbm{1}\{S=s\}\right]} .
\label{eqn:cond-independence}
\end{align}
This yields
\begin{align}
\left|
\E\{f(X;\beta)\mid F=0,S=s\}
-
\E\{f(X;\beta)\mid F=1,S=s\}
\right|
\leq
\epsilon .
\label{eqn:conditional-statistical-parity-score}
\end{align}
Thus, fairness constraints in this framework regulate the counterfactual prediction score under $Q(\delta)$, for example by requiring comparable predicted counterfactual risk across sensitive groups.
\end{example1}


\begin{remark}[Contrasts between intervention regimes]\label{rmk:conditional-effects}
Letting $m_{\delta}(x)=\E\{Y^{Q(\delta)}\mid X=x\}$, for $\delta_1\neq\delta_0$, the conditional-mean contrast
\[
\tau_{\delta_1,\delta_0}(x)
=
m_{\delta_1}(x)-m_{\delta_0}(x)
\]
is identified under the same assumptions used to identify each regime-specific counterfactual mean above. Thus, a finite-dimensional summary may be defined, for a suitable loss $\mathcal L$, through
\[
\underset{\beta\in\R^k}{\operatorname{minimize}}\;
\E\!\left[
\mathcal L\left\{
f(X;\beta),
\tau_{\delta_1,\delta_0}(X)
\right\}
\right]
+\lambda\rho(\beta)
\quad\text{subject to}\quad
\beta\in\mathcal C.
\]
The smoothness conditions required for estimation and inference may depend
on the choice of $\mathcal L$. Alternatively, regime-specific prediction rules
may be estimated separately and compared through
\[
f\{x;\beta^*(\delta_1)\}
-
f\{x;\beta^*(\delta_0)\}.
\]
Neither construction requires identification of the joint distribution
of $\{Y^{Q(\delta_1)},Y^{Q(\delta_0)}\}$.
\end{remark}

\textbf{Challenges.}
A central complication is that \eqref{eqn:target-program} is not a standard stochastic program with objective
$\E\{H(\beta,\xi)\}$ for a known integrand $H$ and observable sample
$\xi_1,\ldots,\xi_n$. Two common approaches in stochastic optimization are sample average approximation and stochastic approximation, which in their standard forms rely on directly evaluable sample objectives or stochastic gradients. In our setting, however, the relevant program components involve counterfactual quantities that are not directly observed and must instead be identified and estimated through nuisance functions. Consequently, direct plug-in implementations may inherit first-order bias from nuisance estimation, compromising root-$n$ inference for the resulting optimizer. This motivates a semiparametric construction in which the counterfactual components of \eqref{eqn:target-program} are replaced by debiased, cross-fitted estimators with second-order nuisance remainders. The resulting approximating program can be solved using standard optimization methods while retaining valid root-$n$ large-sample inference under suitable regularity conditions.

\section{Estimation}

In this section, we develop an estimation strategy for the target parameter defined as the solution to the stochastic optimization program \eqref{eqn:target-program}. Our estimation procedure can be summarized as the following three-step sequencing:
\begin{itemize}[align=left]
    \item [\textbf{Step 1.}] Derive the efficient influence function for each coefficient that requires estimation in \eqref{eqn:target-program}.
    \item [\textbf{Step 2.}] Construct the efficient semiparametric estimators for the coefficients in \eqref{eqn:target-program} based on the efficient influence functions derived in Step 1, and utilize these estimators to formulate the corresponding approximating program for \eqref{eqn:target-program}.
    \item [\textbf{Step 3.}] Solve the resulting deterministic program with a standard optimizer.
\end{itemize}
We provide a more detailed discussion about each step in the proposed strategy. Our goal in Step 1 is to find the efficient influence function (EIF) for each unknown component in our target program \eqref{eqn:target-program}. 
The EIF enables construction of the efficient semiparametric estimator by de-biasing generic plug-in estimators, where we may achieve local minimax lower bounds \citep{bickel1993efficient, van2002semiparametric,tsiatis2006semiparametric,kennedy2016semiparametric}. This also yields desirable properties for our main estimator in Step 3, such as orthogonality and general second-order bias, which allows us to relax nonparametric conditions on nuisance function estimation.

For fixed $\beta$, let
\[
\ell_\beta(X,Y)
=
L\{Y,f(X;\beta)\},
\quad
\mu^L_{a,\beta}(X)
=
\E\{\ell_\beta(X,Y)\mid X,A=a\},
\]
and write $D_\delta(X)
=
\delta\pi(X)+1-\pi(X)$. Applying the incremental-intervention EIF of
\citet[][Theorem~2]{kennedy2019nonparametric} to the transformed
outcome $\ell_\beta(X,Y)$ gives the following uncentered EIF for $\E\left[
L\{Y^{Q(\delta)},f(X;\beta)\}
\right]$:
\begin{align}
\varphi_{\mathcal R}(Z;\eta_{\mathcal R},\delta,\beta)
&=
\frac{
(\delta A+1-A)
\{\ell_\beta(X,Y)-\mu^L_{A,\beta}(X)\}
+
\delta\pi(X)\mu^L_{1,\beta}(X)
+
\{1-\pi(X)\}\mu^L_{0,\beta}(X)
}
{D_\delta(X)}
\nonumber\\
&\quad+
\frac{
\delta
\{\mu^L_{1,\beta}(X)-\mu^L_{0,\beta}(X)\}
\{A-\pi(X)\}
}
{D_\delta(X)^2}.
\label{eqn:general-risk-eif}
\end{align}
where $\eta_{\mathcal R}
=
\{\pi,\mu^L_{0,\beta},\mu^L_{1,\beta}\}$. By construction, $\E\left\{
\varphi_{\mathcal R}(Z;\eta_{\mathcal R},\delta,\beta)
\right\}
=
\E\left[
L\{Y^{Q(\delta)},f(X;\beta)\}
\right]$.

The usual uncentered EIF for $\E\{Y^{Q(\delta)}\}$ is recovered from
\eqref{eqn:general-risk-eif} by taking $\ell_\beta(X,Y)=Y$; we denote
this special case by $\varphi_Q(Z;\eta_Q,\delta)$, where
$\eta_Q=(\mu_0,\mu_1,\pi)$. Consequently, for
any fixed function $\tau:\mathcal X\to\R$, the uncentered EIF for
$\E\{Y^{Q(\delta)}\tau(X)\}$ is
$\varphi_Q(Z;\eta_Q,\delta)\tau(X)$
\citep[][Lemma~A.1]{kim2022doubly}.

With \eqref{eqn:general-risk-eif}, Step~2 constructs a one step estimator by averaging the EIF evaluated at estimated nuisance functions. This can be justified under classical empirical process conditions or, more flexibly, by sample splitting or cross fitting, which allow modern machine learning nuisance estimators \citep{Chernozhukov18}; see also \citet{kennedy2016semiparametric,kennedy2024semiparametric}.

We then define the sample objective for the approximating program. Since $\lambda\rho(\beta)$ is deterministic for fixed $\beta$, it contributes no additional influence function. Replacing $\eta_{\mathcal R}$ by $\widehat\eta_{\mathcal R}$ gives
$$
\widehat\uppsi(\beta)
=
\Pn\left\{
\varphi_{\mathcal R}(Z;\widehat\eta_{\mathcal R},\delta,\beta)
\right\}
+
\lambda\rho(\beta),
$$
which is minimized over the feasible set. When differentiation under the expectation is valid, the gradient of
the sample criterion is obtained by differentiating
\eqref{eqn:general-risk-eif} with respect to $\beta$ and adding
$\lambda\nabla_\beta\rho(\beta)$. Its stochastic influence-function
contribution is
$\nabla_\beta\varphi_{\mathcal R}
(Z;\eta_{\mathcal R},\delta,\beta)$.

For brevity, for fixed $(\beta,\delta)$, write $\mathcal R_\beta(\delta)
=
\E\left[
L\{Y^{Q(\delta)},f(X;\beta)\}
\right]$. Under standard one-step estimation conditions, each stochastic
component of the approximating program is asymptotically linear. In
particular,
\begin{align}
n^{1/2}
\left[
\Pn\left\{
\varphi_{\mathcal R}(Z;\widehat\eta_{\mathcal R},\delta,\beta)
\right\}
-
\mathcal R_\beta(\delta)
\right]
&=
n^{1/2}(\Pn-\Pb)
\left\{
\varphi_{\mathcal R}(Z;\eta_{\mathcal R},\delta,\beta)
\right\}
+o_{\Pb}(1)
\nonumber\\
&\xrightarrow{d}
N\left(
0,
\var\left\{
\varphi_{\mathcal R}(Z;\eta_{\mathcal R},\delta,\beta)
\right\}
\right),
\label{eqn:objective-root-n-CAN}
\end{align}
provided the empirical-process contribution from estimating
$\eta_{\mathcal R}$ and the second-order von Mises remainder are both
$o_{\Pb}(n^{-1/2})$; see, for example,
\citet[][Section~4]{kennedy2024semiparametric}. Since
$\varphi_{\mathcal R}-\mathcal R_\beta(\delta)$ is the efficient
influence function for the loss-based risk, the resulting one-step
estimator is semiparametrically efficient under the usual regularity
conditions.

Finally in Step 3, the approximating program derived in Step 2 (or its equivalent reformulation) can be solved using a standard optimization routine. Our proposed estimator for counterfactual regression is $f(X;\widehat{\beta})$, where $\widehat{\beta}$ is the optimal solution to the approximating program from Step 3.

In what follows, we illustrate the proposed estimation procedure through examples.

\begin{example2}[Cross-entropy loss] \label{example:cross-entropy-loss}
The proposed framework is also applicable to a classification task. For simplicity, assume $\mathcal{Y} = \{0,1\}$. Then we may consider the cross-entropy loss $L(y,\widehat{y})=- y\log \widehat{y} - (1-y) \log (1-\widehat{y})$. For a generalized linear model $f(X;\beta)$, the corresponding risk is defined by $-\E\left\{ Y^{Q(\delta)}\log f(X;\beta) + (1-Y^{Q(\delta)})\log(1-f(X;\beta)) \right\}$, yielding the EIF
\begin{align*}
        -\left\{ \varphi_Q(Z;\eta_Q,\delta) \log f(X;\beta) + (1-\varphi_Q(Z;\eta_Q,\delta))\log(1-f(X;\beta)) \right\}.
\end{align*}
This leads to the following approximating program:
\begin{align*}
    & \underset{\beta \in \R^k}{\text{minimize}} \quad -\Pn\left\{ \varphi_Q(Z;\widehat{\eta}_Q,\delta) \log f(X;\beta) + (1-\varphi_Q(Z;\widehat{\eta}_Q,\delta))\log(1-f(X;\beta)) \right\}.
\end{align*}
\citet{kim2022doubly} studied counterfactual classification within the framework developed in this work, specifically focusing on the case of a deterministic intervention ($\delta = 0$) and deterministic constraints, where $\mathcal{C}$ is known and fixed. 
\end{example2}

\begin{example2}[Mean squared logarithmic loss]
\label{example:mean-squared-logarithmic-loss}
Suppose
\[
\log\{1+f(X;\beta)\}=\beta^\top c(X)
\]
for a known finite-dimensional dictionary $c(X)$. Consider the mean
squared logarithmic loss
$L(y,\widehat y)=\frac12\{\log(1+y)-\log(1+\widehat y)\}^2$
with the $L_2$ penalty \citep{hodson2021mean}. The EIF for the
counterfactual risk is
\[
\frac12\log^2\{1+f(X;\beta)\}
-
\log\{1+f(X;\beta)\}\varphi'_Q(Z;\eta'_Q,\delta)
+
\frac12\varphi''_Q(Z;\eta''_Q,\delta),
\]
where $\varphi'_Q$ and $\varphi''_Q$ are defined as $\varphi_Q$ with
$Y$ replaced by $\log(1+Y)$ and $\log^2(1+Y)$, respectively, with the
corresponding nuisance collections denoted by $\eta'_Q$ and
$\eta''_Q$. Hence,
up to terms independent of $\beta$, the approximating problem is
\[
\underset{\beta\in\R^k}{\operatorname{minimize}}
\quad
\frac12\Pn\!\left[\{\beta^\top c(X)\}^2\right]
-
\Pn\!\left[
\beta^\top c(X)
\varphi'_Q(Z;\widehat\eta'_Q,\delta)
\right]
+
\lambda\|\beta\|_2^2.
\]
Thus, the objective depends on the data through finite-dimensional
stochastic components.
\end{example2}

\begin{example2}[$L_2$ loss with fairness criterion]
\label{example:l2-loss}
Consider the linear aggregation form \eqref{eqn:function-aggregation}
with squared-error loss
$L(y,\widehat y)=\frac12(y-\widehat y)^2$
and an $L_2$ penalty. Up to terms independent of $\beta$, the
approximating objective is
\[
\frac12\beta^\top
\Pn\{b(X)b(X)^\top\}\beta
-
\beta^\top
\Pn\{\varphi_Q(Z;\widehat\eta_Q,\delta)b(X)\}
+
\lambda\|\beta\|_2^2.
\]
Suppose in addition that statistical parity
\eqref{eqn:independence} is imposed with respect to a binary sensitive
attribute $F$. The resulting approximating program is
\begin{align*}
\underset{\beta\in\R^k}{\operatorname{minimize}}
\quad&
\frac12\beta^\top
\Pn\{b(X)b(X)^\top\}\beta
-
\beta^\top
\Pn\{\varphi_Q(Z;\widehat\eta_Q,\delta)b(X)\}
+
\lambda\|\beta\|_2^2
\\
\text{subject to}\quad&
\left|
\Pn\left[
\left\{
\frac{1-F}{\Pn(1-F)}
-
\frac{F}{\Pn(F)}
\right\}
b(X)^\top
\right]\beta
\right|
\leq\epsilon.
\end{align*}
\end{example2}

The preceding examples introduce model classes, losses, and constraints separately. We now give two representative combinations to show how these components define concrete counterfactual regression problems to which the proposed framework can be applied.

\emph{(i) Counterfactual no-show risk under shifted reminder intensity.}
Let $Y\in\{0,1\}$ indicate an appointment no-show and $A\in\{0,1\}$ whether an SMS
reminder was sent, and let $Q(\delta)$ shift the odds of sending reminders relative
to current practice. The target $\E\{Y^{Q(\delta)}\mid X=x\}=\Pb\{Y^{Q(\delta)}=1\mid X=x\}$
is the no-show risk under reminder intensity $\delta$. Since $Y$ is binary, one may use
the cross-entropy loss (Example~\ref{example:cross-entropy-loss}) with a model
$f(x;\beta)\in(0,1)$, or the $L_2$ loss (Example~\ref{example:l2-loss}), and encode
monotonicity of risk in waiting time via linear constraints $C\beta\le d$ as in
Example~\ref{example:constrained-regression}. The resulting estimator is a constrained
counterfactual risk score, not a treatment-allocation rule; for an effect-based ranking, the appropriate target is the identified conditional-mean contrast in Remark~\ref{rmk:conditional-effects}. We revisit this as our case study in Section~\ref{subsec:case-study}.

\emph{(ii) Fair counterfactual risk scoring under an access shift.}
Suppose $Y$ is an adverse outcome, $A$ a treatment or service, and $Q(\delta)$ a
prespecified contraction ($\delta<1$) or expansion ($\delta>1$) of treatment intensity.
A planner facing such a shift may want to identify who is predicted to fare worst under
the new regime, while requiring the resulting counterfactual score to satisfy a parity
constraint across protected groups. The linear ensemble in \eqref{eqn:function-aggregation}, combined with $L_2$ loss (Example~\ref{example:l2-loss}) and statistical parity \eqref{eqn:independence},
gives a score that is both adapted to the shifted regime and constrained to satisfy the
specified fairness requirement. This is different from either fitting a prediction model
under the historical regime or estimating only the marginal mean
$\E\{Y^{Q(\delta)}\}$, neither of which yields a covariate-specific constrained
counterfactual risk score.

\section{Asymptotic Analysis} \label{sec:analysis}

We develop pointwise asymptotic theory for a fixed intervention
parameter $\delta$. To this end, write the target program
\eqref{eqn:target-program} generically as
\begin{equation} \label{eqn:shapiro-true}
\begin{aligned} 
    \underset{\beta \in \R^k}{\text{minimize}} \,\,  \uppsi(\beta)  
    \quad \text{subject to} \quad g_j(\beta) \leq 0, \, j=1,\ldots,r, 
\end{aligned} \tag{$\textsf{P}_{g}$}
\end{equation}
with corresponding approximating program
\begin{equation} \label{eqn:shapiro-approx}
\begin{aligned} 
    \underset{\beta \in \R^k}{\text{minimize}} \,\,  \widehat{\uppsi}(\beta)  
    \quad \text{subject to} \quad \widehat{g}_j(\beta) \leq 0, \, j=1,\ldots,r.
\end{aligned} \tag{$\widehat{\textsf{P}}_{g}$}
\end{equation}
Here,
\[
\uppsi(\beta)
=
\E\left\{
\varphi_{\mathcal R}
(Z;\eta_{\mathcal R},\delta,\beta)
\right\}
+
\lambda\rho(\beta),
\quad
\widehat{\uppsi}(\beta)
=
\Pn\left\{
\varphi_{\mathcal R}
(Z;\widehat\eta_{\mathcal R},\delta,\beta)
\right\}
+
\lambda\rho(\beta),
\]
while $\widehat g_j$ denotes an estimator of $g_j$ when the constraint
is data dependent.

Classical stochastic-programming asymptotics typically rely on
localization of the primal--dual solution, for example,
\[
\widehat\beta\xrightarrow{p}\beta^*,
\quad
\widehat\gamma\xrightarrow{p}\gamma^*,
\]
together with regularity of the solution map
\citep[e.g.,][]{dupacova1988asymptotic,shapiro1993asymptotic,
shapiro2000statistical}. Rather than imposing joint primal--dual consistency as a common
high-level condition, we consider two program classes for which the
joint consistency can be established from class-specific regularity
conditions. Section~\ref{subsec:smooth-with-fixed-set} considers smooth objectives with a fixed
feasible set, where primal consistency follows from empirical criterion
regularity and dual consistency from local regularity of the
Karush--Kuhn--Tucker (KKT) system. Section~\ref{subsec:estimated-linear-constraints}
considers finite-dimensional stochastic components with estimated
linear constraints, where the joint consistency follows from convergence of the
estimated components and local regularity of the population KKT system. The resulting semiparametric perturbations are then propagated through the corresponding solution maps.

\subsection{Smooth objectives with a fixed feasible set} \label{subsec:smooth-with-fixed-set}

First, we consider the case where both \ref{eqn:shapiro-true} and \ref{eqn:shapiro-approx} share a common set of deterministic
constraints $\mathcal{C} = \{\beta : g_j(\beta) \leq 0, j=1,\ldots,r\}$:
\begin{align} 
    & \underset{\beta \in \R^k}{\text{minimize}} \,\,  \uppsi(\beta)  
    \quad \text{subject to} \quad g_j(\beta) \leq 0, \, j=1,\ldots,r, \tag{$\textsf{P}_{f}$} \label{eqn:smooth-fixed-true} \\
    &\underset{\beta \in \R^k}{\text{minimize}} \,\,  \widehat{\uppsi}(\beta)  
    \quad \text{subject to} \quad g_j(\beta) \leq 0, \, j=1,\ldots,r. \tag{$\widehat{\textsf{P}}_{f}$} \label{eqn:smooth-fixed-approx}
\end{align} 
This setting accommodates deterministic smooth constraints, such as linear inequality, shape, or calibration restrictions on $\beta$, but not constraints whose coefficients or defining functions must themselves be estimated from the data.
For any feasible point $\bar{\beta}\in\mathcal C$ in
\eqref{eqn:smooth-fixed-true}, define the \emph{active index set} by
\[
J_0(\bar{\beta})
=
\{1\leq j\leq r:g_j(\bar{\beta})=0\}.
\]
We now introduce some standard regularity conditions for our analysis. 
\begin{definition}[LICQ]
{Linear independence constraint qualification} (LICQ) is satisfied at $\bar{\beta} \in \mathcal{C}$, if the vectors $\nabla_\beta g_j(\bar{\beta})$, $j \in J_0(\bar{\beta})$ are linearly independent. 
\end{definition}
\begin{definition}[SC]
Suppose $\bar\beta$ is feasible and $\bar\gamma$ is a multiplier vector
satisfying the KKT conditions at $\bar\beta$. Strict complementarity
(SC) holds if
\[
\bar\gamma_j>0,
\quad
j\in J_0(\bar\beta).
\]
\end{definition}
LICQ is a standard constraint qualification ensuring the validity of the
first-order KKT necessary conditions at a local optimum. SC requires each
active inequality constraint to have a strictly positive multiplier,
while inactive constraints have zero multipliers by complementary
slackness. Both conditions are standard in nonlinear and parametric
optimization \citep[e.g.,][]{still2018lectures}. We require LICQ and SC at the target optimizer
$\beta^*$.
\begin{assumption}\label{assumption:LICQ-SC}
LICQ and SC hold at the target optimizer $\beta^*$ with associated
KKT multipliers $\gamma^*$.
\end{assumption}
If we assume that the optimal solution $\beta^*$ is unique and LICQ holds at $\beta^*$, then the corresponding multipliers $\gamma^*$ are determined uniquely \citep{wachsmuth2013licq}. We next impose a standard second-order regularity condition.

\begin{assumption}
\label{assumption:quadratic-growth}
The constraint functions $g_j$, $j=1,\ldots,r$, are continuous at
$\beta^*$. In addition, $\uppsi$ and $g_j$, $j\in J_0(\beta^*)$,
are twice continuously differentiable in a neighborhood of
$\beta^*$. Let
\[
B_{ac}
=
\left(
\nabla_\beta g_j(\beta^*)^\top
\right)_{j\in J_0(\beta^*)},
\]
\[
H(\beta^*)
=
\nabla_{\beta}^2\uppsi(\beta^*)
+
\sum_{j\in J_0(\beta^*)}
\gamma_j^*
\nabla_{\beta}^2g_j(\beta^*).
\]
Then, there exists a constant $\kappa>0$ such that
\[
v^\top H(\beta^*)v
\geq
\kappa\|v\|_2^2
\quad
\text{for every }v\text{ satisfying }B_{ac}v=0.
\]
\end{assumption}

Under LICQ and SC,
Assumption~\ref{assumption:quadratic-growth} is the standard
second-order sufficient condition on the tangent space of the active
constraints. It implies local quadratic growth and, together with LICQ,
nonsingularity of the reduced KKT matrix
\[
J_f(\beta^*)
=
\begin{pmatrix}
H(\beta^*) & B_{ac}^\top\\
B_{ac} & 0
\end{pmatrix}.
\]

We also require a central limit theorem for the gradient perturbation $\nabla_\beta \widehat{\uppsi}(\beta^*)-\nabla_\beta \uppsi(\beta^*)$.
\begin{assumption} \label{assumption:grad-rootn-CLT-fixed}
We assume that the gradient estimator $\nabla_{\beta}\widehat{\uppsi}(\beta)$ satisfies, at $\beta=\beta^*$,
\begin{align*}
n^{1/2}\Big\{
\nabla_{\beta}\widehat{\uppsi}(\beta^*)-\nabla_{\beta}\uppsi(\beta^*)
\Big\}
\xrightarrow{d}
N\!\left(0,\ \Sigma_{\nabla}(\beta^*)\right), \quad \text{where} \quad \Sigma_{\nabla}(\beta^*)
=
\var\!\left\{
\nabla_\beta \varphi_{\mathcal R}(Z;\eta_{\mathcal R},\delta,\beta^*)
\right\}.
\end{align*}
\end{assumption}

Assumption~\ref{assumption:grad-rootn-CLT-fixed} can be verified under differentiated versions of the
one-step regularity conditions underlying \eqref{eqn:objective-root-n-CAN},
including interchange of differentiation and expectation and a
root-$n$ negligible nuisance remainder for the gradient; sufficient
conditions are given in Appendix \ref{appsec:regularity-grad-rootn-CLT-fixed}.

Next, we impose empirical criterion regularity conditions that localize the global sample
optimizer and stabilize the empirical KKT Jacobian.

\begin{assumption}
\label{assumption:localization-fixed}
For every $\epsilon>0$,
\[
\inf_{\substack{\beta\in\mathcal C\\
\|\beta-\beta^*\|_2\geq\epsilon}}
\left\{
\uppsi(\beta)-\uppsi(\beta^*)
\right\}
>0 \quad \text{ and } \quad
\sup_{\beta\in\mathcal C}
\left|
\widehat{\uppsi}(\beta)-\uppsi(\beta)
\right|
=
o_{\Pb}(1).
\]
Moreover, there exists a neighborhood $\mathcal N$ of $\beta^*$ such
that, with probability tending to one,
$\widehat{\uppsi}$ is twice continuously differentiable on $\mathcal N$
and
\[
\sup_{\beta\in\mathcal N}
\left\|
\nabla_{\beta}^2\widehat{\uppsi}(\beta)
-
\nabla_{\beta}^2\uppsi(\beta)
\right\|_2
=
o_{\Pb}(1).
\]
\end{assumption}

Assumption~\ref{assumption:localization-fixed} uses standard
extremum-estimation conditions: uniform criterion convergence and
separation yield consistency, while local Hessian convergence
supports the subsequent linearization
\citep[e.g.,][Section~5.2]{van2000asymptotic}. Because the feasible set is fixed, no constraint-convergence condition is required; together with the local smoothness conditions, LICQ and SC
then yield dual consistency and active-set stability. For noncompact feasible sets, uniform criterion convergence may instead be imposed on an appropriate compact
sublevel set.
 
Under these conditions, the closed-form limiting distribution of $\widehat{\beta}$ is stated below.

\begin{theorem}
\label{thm:asymptotics-fixed}
Assume that \eqref{eqn:smooth-fixed-true} has a unique global optimizer
$\beta^*$, and let $\widehat\beta$ be a measurable global optimizer of
\eqref{eqn:smooth-fixed-approx}. Suppose
Assumptions~\ref{assumption:LICQ-SC},
\ref{assumption:quadratic-growth},
\ref{assumption:grad-rootn-CLT-fixed}, and
\ref{assumption:localization-fixed} hold. Then
\[
\widehat\beta\xrightarrow{p}\beta^*.
\]
Moreover, with probability tending to one,
$\widehat\beta$ has a unique KKT multiplier vector
$\widehat\gamma$, the sample active set equals $J_0(\beta^*)$, and
\[
\widehat\gamma_{ac}\xrightarrow{p}\gamma^*_{ac}.
\]
Finally,
\[
n^{1/2}
\begin{pmatrix}
\widehat\beta-\beta^*\\
\widehat\gamma_{ac}-\gamma^*_{ac}
\end{pmatrix}
\xrightarrow{d}
-
J_f(\beta^*)^{-1}
\begin{pmatrix}
W\\
0
\end{pmatrix},
\]
where $W\sim N\!\left\{0,\Sigma_\nabla(\beta^*)\right\}$. Consequently,
\[
n^{1/2}(\widehat\beta-\beta^*)
\xrightarrow{d}
-
\begin{pmatrix}I_k&0\end{pmatrix}
J_f(\beta^*)^{-1}
\begin{pmatrix}
W\\
0
\end{pmatrix}.
\]
\end{theorem}
Proofs of all theoretical results are provided in Appendix~\ref{appsec:proofs}.
Theorem~\ref{thm:asymptotics-fixed} establishes the first-order limit
without imposing joint primal--dual consistency directly. We next
specialize this result to the cross-entropy loss.

\begin{corollary}\label{cor:cross-entropy}
Consider the cross-entropy loss in
Example~\ref{example:cross-entropy-loss}. Suppose the conditions of
Theorem~\ref{thm:asymptotics-fixed}, except
Assumption~\ref{assumption:grad-rootn-CLT-fixed}, hold. Assume further
that $f$ is twice continuously differentiable with respect to $\beta$,
that there exists $\underline c\in(0,1/2)$ such that $\underline c
\leq
f(X;\beta)
\leq
1-\underline c$ almost surely for all $\beta$ in a neighborhood of $\beta^*$. If
\begin{align*}
\max_{a}
\|\widehat\mu_a-\mu_a\|_{2,\Pb}
+
\|\widehat\pi-\pi\|_{2,\Pb}
&=
o_{\Pb}(1),\\
\|\widehat\pi-\pi\|_{2,\Pb}
\left\{
\max_{a}
\|\widehat\mu_a-\mu_a\|_{2,\Pb}
+
\|\widehat\pi-\pi\|_{2,\Pb}
\right\}
&=
o_{\Pb}(n^{-1/2}).
\end{align*}
then, under the sufficient conditions in
Section~\ref{appsec:regularity-grad-rootn-CLT-fixed},
Assumption~\ref{assumption:grad-rootn-CLT-fixed} holds. Consequently,
$\widehat\beta$ is $\sqrt n$-consistent and asymptotically normal for
$\beta^*$.
\end{corollary}

\subsection{Finite-dimensional stochastic components with estimated linear constraints}\label{subsec:estimated-linear-constraints}

We next consider an alternative structure for
\eqref{eqn:shapiro-true}, in which the counterfactual risk and possibly
the constraints depend on a finite-dimensional vector of stochastic
components. This provides a separate route to consistency and
asymptotic normality from that of
Section~\ref{subsec:smooth-with-fixed-set}.

Let $\mathbbm T$ be a finite-dimensional Euclidean set, and let
$T=T(\Pb)\in\mathbbm T$ denote a statistical functional of the
observed data distribution. Let $h:\R^k\times\mathbbm{T}\to\R$ be twice continuously differentiable. We also allow the linear-constraint coefficients $C\in\R^{r\times k}$ and $d\in\R^r$ to depend on $\Pb$, so that $(T,C,d)$ are unknown and must be estimated. We assume, however, that the functional form of $h$ is known and deterministic, and that all dependence on $\Pb$ enters only through $(T,C,d)$; that is, the stochastic components are separable from $h$. In this formulation, $T(\Pb)$ represents identified population components of interest. 

A sufficient structure for this formulation is that the counterfactual risk can be reduced to finitely many identified population components. Specifically, suppose that, up to terms not depending on $\beta$,
$$
\mathcal R\{f(X;\beta),Y^{Q(\delta)}\}+\lambda\rho(\beta)
=
h(\beta,T),
\quad
T=(T_1,\ldots,T_q)\in\mathbbm T,
$$
where each $T_\ell=T_\ell(\Pb)$ is an identified scalar functional of the observed data distribution. A typical case is
$$
h(\beta,T)
=
h_0(\beta)
+
\sum_{\ell=1}^q h_\ell(\beta)T_\ell,
$$
where $h_0,h_1,\ldots,h_q$ are known once the loss, model class, and penalty are chosen. Thus, $T$ collects the population moments needed to evaluate the risk, while $h$ deterministically combines those moments with $\beta$. Estimated linear constraints, such as fairness, calibration, or balance restrictions, are handled similarly through finite dimensional coefficients $C=C(\Pb)$ and $d=d(\Pb)$. Hence this section covers objectives and constraints admitting finite dimensional reductions; arbitrary nonlinear or nonsmooth data adaptive constraints require additional regularity. Corollaries~\ref{cor:1} and~\ref{cor:2} provide concrete examples of this structure.

Our goal is to estimate the optimal solution of
\begin{equation}
\label{eqn:general-separable-case-true}
\begin{aligned}    
    \underset{\beta \in \R^k}{\text{minimize}} \,\,  h(\beta, T)  
    \quad \text{subject to} \quad C\beta \leq d,
\end{aligned}       \tag{$\textsf{P}_{sl}$}
\end{equation}
where we write $T(\Pb) \equiv T$. Since the true program \eqref{eqn:general-separable-case-true} is not directly solvable, we compute an optimal solution of the following approximating program:
\begin{equation}
\label{eqn:general-separable-case-approx}
\begin{aligned}    
    \underset{\beta \in \R^k}{\text{minimize}} \,\,  h(\beta, \widehat{T})  
    \quad \text{subject to} \quad \widehat{C}\beta \leq \widehat{d}.
\end{aligned}       \tag{$\widehat{\textsf{P}}_{sl}$}  
\end{equation}

This setup allows the feasible set to vary with $\Pb$: the constraint pairs $(C,d)$ and $(\widehat C,\widehat d)$ need not coincide, so \eqref{eqn:general-separable-case-true} and \eqref{eqn:general-separable-case-approx} generally have different, data-adaptive feasible regions, each given by linear inequalities. It specifically accommodates estimated linear constraints of the form $\widehat C\beta\le \widehat d$; the fixed shape constraints in
Example~\ref{example:constrained-regression} and the estimated
fairness constraints in Example~\ref{example:fairness} both fit
this formulation. A canonical choice of $h$ is the $L_2$ risk for the linear aggregation model \eqref{eqn:function-aggregation}, as in Example~\ref{example:l2-loss}. The theory does not cover arbitrary nonlinear or nonsmooth data-adaptive constraints without additional regularity conditions.

Fix a target optimizer
$\beta^*\in\sol(\text{\ref{eqn:general-separable-case-true}})$.
For this linear-constraint program, write
\[
J_0
=
J_0(\beta^*)
=
\{1\leq j\leq r:C_j\beta^*=d_j\},
\quad
m=|J_0|.
\]
Let $C_{ac}\in\R^{m\times k}$ and $d_{ac}\in\R^m$ denote the rows
of $C$ and $d$ indexed by $J_0$. For constraint-indexed quantities,
the subscript $ac$ denotes restriction to these same indices; thus
$\widehat C_{ac}$ and $\widehat d_{ac}$ denote the corresponding
estimated quantities.

We impose local regularity of the population KKT system.

\begin{assumption}
\label{assumption:strong-reg-sl}
The function $h$ is twice continuously differentiable in a neighborhood
of $(\beta^*,T)$. The pair $(\beta^*,\gamma^*_{ac})$ satisfies the
active-set KKT conditions, $C_{ac}$ has full row rank, and
$\gamma^*_{ac}>0$ componentwise. Moreover, there exists $\kappa>0$
such that
\[
v^\top
\nabla_{\beta}^2h(\beta^*,T)v
\geq
\kappa\|v\|_2^2
\quad
\text{for every }v\text{ satisfying }C_{ac}v=0.
\]
\end{assumption}

Under Assumption~\ref{assumption:strong-reg-sl}, the reduced KKT matrix
\[
M
=
\begin{pmatrix}
\nabla_{\beta}^2h(\beta^*,T) & C_{ac}^\top\\
C_{ac} & 0
\end{pmatrix}
\]
is nonsingular. We also require the following mild consistency condition, with no requirement on rates of convergence.
\begin{assumption} \label{assumption:consistency}
    $\max\left\{ \Vert \widehat{T} - T \Vert_2, \Vert \widehat{C} - C \Vert_F, \Vert \widehat{d} - d \Vert_2 \right\} = o_{\Pb}(1)$.
\end{assumption}
The local stability of the estimator near $\beta^*$ is formalized in the next theorem.
\begin{theorem} 
\label{thm:rates}
Suppose Assumptions~\ref{assumption:strong-reg-sl} and
\ref{assumption:consistency} hold. Then, with probability tending to one, there exists a unique KKT pair
$(\widehat\beta,\widehat\gamma_{ac})$ in a neighborhood of
$(\beta^*,\gamma^*_{ac})$, with $\widehat\beta$ a strict local
minimizer of the sample program, and
\[
\left\|
\begin{pmatrix}
\widehat\beta-\beta^*\\
\widehat\gamma_{ac}-\gamma^*_{ac}
\end{pmatrix}
\right\|_2
=
O_{\Pb}\!\left(
\max\left\{
\|\widehat T-T\|_2,\,
\|\widehat C-C\|_F,\,
\|\widehat d-d\|_2
\right\}
\right).
\]
If $h(\cdot,T')$ is strongly convex on $\R^k$ for every $T'$ in a
neighborhood of $T$, this local solution is the unique global optimizer
of \eqref{eqn:general-separable-case-approx}.
\end{theorem}

Theorem~\ref{thm:rates} exploits the finite-dimensional linear-constraint
representation to transfer consistency of $(\widehat T,\widehat C,
\widehat d)$ to the local primal--dual solution through the KKT map.
Thus, primal--dual consistency need not be imposed directly.

For the root-$n$ expansion below, we impose root-$n$ consistency of the
estimated finite-dimensional program components.

\begin{assumption}\label{assumption:root-n-rates} $\max\left\{
\|\widehat T-T\|_2,\,
\|\widehat C-C\|_F,\,
\|\widehat d-d\|_2
\right\}
=
O_{\Pb}(n^{-1/2}).$
\end{assumption}

Assumption~\ref{assumption:root-n-rates} does not require the underlying
nuisance functions to be estimated at parametric rates. In our
semiparametric construction, $\widehat T$ admits a root-$n$ expansion
when the relevant second-order products of nuisance estimation errors are
$o_{\Pb}(n^{-1/2})$; see
Corollaries~\ref{cor:1}--\ref{cor:2}. Likewise, estimators of $C$ and
$d$ based on empirical moments or smooth functions thereof typically
admit root-$n$ expansions under standard regularity conditions.

The linear-constraint structure makes the relevant program perturbation
finite dimensional: estimation of $(T,C,d)$ perturbs the active
stationarity equation and the affine constraint residuals directly.
Thus, rather than imposing a first-order expansion for the optimizer
itself, we impose one for these KKT perturbations.

\begin{assumption}[First-order KKT perturbation]
\label{assumption:kkt-convergence}
Define
\[
\Xi_n
=
n^{1/2}
\begin{pmatrix}
\nabla_\beta h(\beta^*,\widehat T)
-
\nabla_\beta h(\beta^*,T)
+
(\widehat C_{ac}-C_{ac})^\top\gamma^*_{ac}
\\[3pt]
(\widehat C_{ac}-C_{ac})\beta^*
-
(\widehat d_{ac}-d_{ac})
\end{pmatrix}
\in\R^{k+m}.
\]
We assume
\[
\Xi_n\xrightarrow{d}\Xi
\]
for some random vector $\Xi\in\R^{k+m}$.
\end{assumption}

Assumption~\ref{assumption:kkt-convergence} concerns the joint
first-order perturbation of the stationarity equation and the active
constraint residuals, rather than convergence of the KKT multipliers.
It holds when the relevant components of
$(\widehat T,\widehat C,\widehat d)$ admit a joint asymptotically linear
representation. 

The next theorem gives a closed-form first-order expansion for the
estimator $\widehat\beta$ characterized in Theorem~\ref{thm:rates}.

\begin{theorem}
\label{thm:asymptotics}
Suppose Assumptions~\ref{assumption:strong-reg-sl},
\ref{assumption:root-n-rates}, and
\ref{assumption:kkt-convergence} hold. Let
$(\widehat\beta,\widehat\gamma_{ac})$ denote the local solution from
Theorem~\ref{thm:rates}. Then
\[
n^{1/2}
\begin{pmatrix}
\widehat\beta-\beta^*\\
\widehat\gamma_{ac}-\gamma^*_{ac}
\end{pmatrix}
=
-
M^{-1}\Xi_n
+
o_{\Pb}(1).
\]
Consequently,
\[
n^{1/2}(\widehat\beta-\beta^*)
=
-
\begin{pmatrix}I_k&0\end{pmatrix}
M^{-1}\Xi_n
+
o_{\Pb}(1),
\]
and hence
\[
n^{1/2}(\widehat\beta-\beta^*)
\xrightarrow{d}
-
\begin{pmatrix}I_k&0\end{pmatrix}
M^{-1}\Xi.
\]
If $\Xi$ is mean-zero Gaussian, the limiting distribution is
multivariate normal.
\end{theorem}

Theorem~\ref{thm:asymptotics} propagates the joint semiparametric
perturbation of the objective and active constraints through the locally
differentiable KKT solution map. It provides a basis for Wald or
bootstrap inference whenever the distribution of $\Xi$ is consistently
estimated. We apply Theorem~\ref{thm:asymptotics} to
Examples~\ref{example:mean-squared-logarithmic-loss} and
\ref{example:l2-loss}, yielding the following corollaries.

\begin{corollary}\label{cor:1}
Consider the mean squared logarithmic loss with the $L_2$ penalty term
as in Example~\ref{example:mean-squared-logarithmic-loss}. Suppose the
local KKT regularity conditions of
Theorem~\ref{thm:asymptotics} hold, and the finite-dimensional
empirical and influence-function components entering $T$ have finite
second moments. Let $\mu'_a(X)
=
\E\{\log(1+Y)\mid X,A=a\}, a\in\{0,1\}$. Assume that
\begin{align*}
\max_{a}
\|\widehat\mu'_a-\mu'_a\|_{2,\Pb}
+
\|\widehat\pi-\pi\|_{2,\Pb}
&=
o_{\Pb}(1),\\
\|\widehat\pi-\pi\|_{2,\Pb}
\left\{
\|\widehat\pi-\pi\|_{2,\Pb}
+
\max_{a}
\|\widehat\mu'_a-\mu'_a\|_{2,\Pb}
\right\}
&=
o_{\Pb}(n^{-1/2}).
\end{align*}
Then the conditions of Theorem~\ref{thm:asymptotics} are satisfied, so
$\widehat\beta$ is $\sqrt n$-consistent and asymptotically normal for
$\beta^*$.
\end{corollary}

\begin{corollary}\label{cor:2}
Consider the $L_2$ loss with the $L_2$ penalty, the aggregated
predictors \eqref{eqn:function-aggregation}, and statistical parity as
in Example~\ref{example:l2-loss}. Suppose Assumption~\ref{assumption:strong-reg-sl} holds,
$0<\Pb(F=1)<1$, and the finite-dimensional empirical and
influence-function components entering $(T,C,d)$ have finite second
moments. Assume that
\begin{align*}
\max_{a}
\|\widehat\mu_a-\mu_a\|_{2,\Pb}
+
\|\widehat\pi-\pi\|_{2,\Pb}
&=
o_{\Pb}(1),\\
\|\widehat\pi-\pi\|_{2,\Pb}
\left\{
\|\widehat\pi-\pi\|_{2,\Pb}
+
\max_{a}
\|\widehat\mu_a-\mu_a\|_{2,\Pb}
\right\}
&=
o_{\Pb}(n^{-1/2}).
\end{align*}
Then the conditions of Theorem~\ref{thm:asymptotics} are satisfied, so
$\widehat\beta$ is $\sqrt n$-consistent and asymptotically normal for
$\beta^*$.
\end{corollary}

The rate conditions in Corollaries~\ref{cor:1}--\ref{cor:2} are
second-order product conditions for incremental interventions, applied
to $\log(1+Y)$ in Corollary~\ref{cor:1} and to $Y$ in
Corollary~\ref{cor:2}. In the latter, the statistical-parity constraint
is based on finite-dimensional empirical moments and therefore admits
standard $\sqrt{n}$ behavior. The same argument extends to other smooth
moment-based linear constraints, whereas nonsmooth or nonlinear
data-adaptive constraints require additional regularity.

\section{Uniform Inference over $\delta$}
\label{sec:uniform-delta}

Section~4 established pointwise inference for a fixed $\delta$. We now
extend these results uniformly over a compact range
$\Delta=[\delta_{\min},\delta_{\max}]\subset(0,\infty)$.
Section~\ref{subsec:uniform-fixed} treats smooth objectives with a fixed
feasible set, while Section~\ref{subsec:uniform-stochastic-linear}
treats finite-dimensional components with estimated linear constraints,
mirroring the two pointwise settings in Section~4. The assumptions below
are corresponding uniform versions of the pointwise conditions.

For each $\delta\in\Delta$, let $\beta^*(\delta)$ denote the population
optimizer. Under the stated conditions, $\widehat\beta(\delta)$ denotes
the unique local optimizer in a common neighborhood of
$\beta^*(\delta)$; for convex programs with a unique global optimizer,
the two coincide. Section~\ref{subsec:bands} then uses the resulting
uniform first-order expansions to construct simultaneous confidence
bands for coordinates of $\beta^*(\cdot)$ and for the prediction path
$\delta\mapsto f\{x;\beta^*(\delta)\}$ at prespecified covariate
profiles $x$.

\subsection{Fixed feasible set}
\label{subsec:uniform-fixed}

We first consider the fixed feasible set setting of Section~\ref{subsec:smooth-with-fixed-set}, where $\widehat g_j=g_j$ for $j=1,\ldots,r$. For brevity, define
\begin{align*}
\Psi(\delta,\beta)
&=
\Pb\left\{
\varphi_{\mathcal R}(Z;\eta_{\mathcal R},\delta,\beta)
\right\},
&
\uppsi_\delta(\beta)
&=
\Psi(\delta,\beta)+\lambda\rho(\beta),
\\
\widehat\Psi(\delta,\beta)
&=
\Pn\left\{
\varphi_{\mathcal R}(Z;\widehat\eta_{\mathcal R},\delta,\beta)
\right\},
&
\widehat\uppsi_\delta(\beta)
&=
\widehat\Psi(\delta,\beta)+\lambda\rho(\beta).
\end{align*}
Let
$$
G(z;\eta_{\mathcal R},\delta,\beta)
=
\nabla_\beta\varphi_{\mathcal R}(z;\eta_{\mathcal R},\delta,\beta),
\quad
\widehat G(z;\delta,\beta)
=
\nabla_\beta\varphi_{\mathcal R}(z;\widehat\eta_{\mathcal R},\delta,\beta).
$$
Define the oracle centered gradient process
$$
\mathbb U_n(\delta)
=
n^{1/2}(\Pn-\Pb)
\left\{
G(Z;\eta_{\mathcal R},\delta,\beta^*(\delta))
\right\}
\in\R^k .
$$

We impose two uniform conditions: a uniform gradient expansion and
uniform regularity of the KKT system along the solution path.

\begin{assumption}
\label{ass:unif-grad-fclt}
The process $\mathbb U_n$ satisfies
\[
\mathbb U_n \rightsquigarrow \mathbb U
\quad\text{in}\quad
\ell^\infty(\Delta)^k,
\]
where $\mathbb U$ is a mean-zero tight Gaussian process with covariance
induced by
$G\{Z;\eta_{\mathcal R},\delta,\beta^*(\delta)\}$. Moreover,
\[
\sup_{\delta\in\Delta}
\left\|
n^{1/2}
\left[
\nabla_\beta\widehat\uppsi_\delta\{\beta^*(\delta)\}
-
\nabla_\beta\uppsi_\delta\{\beta^*(\delta)\}
\right]
-
\mathbb U_n(\delta)
\right\|_2
=
o_{\Pb}(1).
\]
\end{assumption}
Assumption~\ref{ass:unif-grad-fclt} is the uniform counterpart of
Assumption~\ref{assumption:grad-rootn-CLT-fixed}, strengthening the
pointwise gradient CLT to a functional CLT and uniform first-order
expansion over $\delta\in\Delta$.

\begin{assumption}
\label{ass:unif-strong-reg}
Let
$B_f(\delta)=\bigl(\nabla_\beta g_j\{\beta^*(\delta)\}^\top\bigr)_{j\in J_0}$
and
\[
H_f(\delta)
=
\nabla_\beta^2\uppsi_\delta\{\beta^*(\delta)\}
+
\sum_{j\in J_0}
\gamma_j^*(\delta)\nabla_\beta^2g_j\{\beta^*(\delta)\}.
\]
The following conditions hold:
\begin{enumerate}[label=(\roman*),leftmargin=2em]
\item The solution and multiplier paths are continuous on $\Delta$, the
active set is constant, $J_0(\delta)=J_0$, with $m=|J_0|$, and uniform
LICQ and SC hold:
\[
\inf_{\delta\in\Delta}
\lambda_{\min}\{B_f(\delta)B_f(\delta)^\top\}>0,
\quad
\inf_{\delta\in\Delta}
\min_{j\in J_0}\gamma_j^*(\delta)>0,
\]
with these conditions vacuous when $m=0$.

\item For some $\kappa>0$,
$v^\top H_f(\delta)v\geq\kappa\|v\|_2^2$ whenever
$B_f(\delta)v=0$, uniformly over $\delta\in\Delta$.

\item The functions $\uppsi_\delta$, $g_j$, and
$\widehat\uppsi_\delta$ are twice continuously differentiable on a
common neighborhood of the solution path, with probability tending to
one for $\widehat\uppsi_\delta$, and, for some $r>0$,
\[
\sup_{\delta\in\Delta}
\sup_{\|\beta-\beta^*(\delta)\|_2\leq r}
\left\|
\nabla_\beta^2\widehat\uppsi_\delta(\beta)
-
\nabla_\beta^2\uppsi_\delta(\beta)
\right\|_2
=
o_{\Pb}(1).
\]
\end{enumerate}
\end{assumption}

Conditions (i)-(ii) of
Assumption~\ref{ass:unif-strong-reg} imply uniform nonsingularity of the
reduced population KKT matrix
\[
J_f(\delta)
=
\begin{pmatrix}
H_f(\delta) & B_f(\delta)^\top\\
B_f(\delta) & 0
\end{pmatrix}.
\]
They are uniform counterparts of standard regularity conditions in
smooth parametric optimization with a stable active set
\citep{shapiro1990differential,shapiro1993asymptotic,
shapiro2000statistical}. Condition~(iii) transfers this local regularity
to the sample KKT system. Together with
Assumption~\ref{ass:unif-grad-fclt}, they yield the uniform localization
and first-order expansion below.

Now, define
\[
\mathsf A_f(\delta)
=
\begin{pmatrix}
I_k&0
\end{pmatrix}
J_f(\delta)^{-1}
\begin{pmatrix}
I_k\\
0
\end{pmatrix}.
\]
The next theorem establishes a uniform first-order asymptotic expansion for the fixed feasible set setting.
\begin{theorem}
\label{thm:unif-linearization}
Suppose Assumptions~\ref{ass:unif-grad-fclt} and
\ref{ass:unif-strong-reg} hold. Then, with probability tending to one,
there exists a unique strict local optimizer $\widehat\beta(\delta)$
in a common neighborhood of $\beta^*(\delta)$ for every
$\delta\in\Delta$, with active set $J_0$ and multiplier
$\widehat\gamma_{ac}(\delta)$, such that
\[
\sup_{\delta\in\Delta}
\left\|
\begin{pmatrix}
\widehat\beta(\delta)-\beta^*(\delta)\\
\widehat\gamma_{ac}(\delta)-\gamma^*_{ac}(\delta)
\end{pmatrix}
\right\|_2
=
O_{\Pb}(n^{-1/2}).
\]
Moreover, in $\ell^\infty(\Delta)^{k+m}$,
\[
n^{1/2}
\begin{pmatrix}
\widehat\beta(\delta)-\beta^*(\delta)\\
\widehat\gamma_{ac}(\delta)-\gamma^*_{ac}(\delta)
\end{pmatrix}
=
-
J_f(\delta)^{-1}
\begin{pmatrix}
\mathbb U_n(\delta)\\
0
\end{pmatrix}
+
o_{\Pb}(1).
\]
Consequently, in $\ell^\infty(\Delta)^k$,
\[
n^{1/2}
\{\widehat\beta(\delta)-\beta^*(\delta)\}
=
-\mathsf A_f(\delta)\mathbb U_n(\delta)
+
o_{\Pb}(1).
\]
\end{theorem}
Unlike the pointwise expansions in Section~4, this result holds uniformly over all $\delta\in\Delta$, and therefore supports simultaneous inference for the entire intervention-response curve.

\subsection{Separable stochastic components and linear constraints}
\label{subsec:uniform-stochastic-linear}

We next consider the setting of
Section~\ref{subsec:estimated-linear-constraints}, now indexed by
$\delta\in\Delta$. For each $\delta$, consider
\[
\underset{\beta\in\R^k}{\operatorname{minimize}}
\quad
h\{\beta,T(\delta)\}
\quad
\text{subject to}
\quad
C(\delta)\beta\le d(\delta),
\]
with sample analogue
\[
\underset{\beta\in\R^k}{\operatorname{minimize}}
\quad
h\{\beta,\widehat T(\delta)\}
\quad
\text{subject to}
\quad
\widehat C(\delta)\beta\le \widehat d(\delta).
\]
Let $\beta^*(\delta)$ and $\gamma^*_{ac}(\delta)$ denote the population
optimizer and active multiplier paths, and let $J_0$ denote the common
active set over $\Delta$, as specified below in
Assumption~\ref{ass:unif-sl-strong-reg}, with $m=|J_0|$. Define
$$
\mathbb V_n(\delta)
=
n^{1/2}
\begin{pmatrix}
\nabla_\beta h\{\beta^*(\delta),\widehat T(\delta)\}
-
\nabla_\beta h\{\beta^*(\delta),T(\delta)\}
+
\left\{\widehat C_{ac}(\delta)-C_{ac}(\delta)\right\}^\top
\gamma^*_{ac}(\delta)
\\[1mm]
\left\{\widehat C_{ac}(\delta)-C_{ac}(\delta)\right\}\beta^*(\delta)
-
\left\{\widehat d_{ac}(\delta)-d_{ac}(\delta)\right\}
\end{pmatrix}
\in\R^{k+m}.
$$

As in the previous subsection, we impose two uniform conditions for this stochastic linear-constraint setting: one on the joint perturbation process $\mathbb V_n(\cdot)$ and one on the uniform KKT regularity of the active constraint system.
\begin{assumption}
\label{ass:unif-sl-process}
The process $\mathbb V_n(\cdot)$ satisfies
\[
\mathbb V_n\rightsquigarrow \mathbb V
\quad\text{in}\quad
\ell^\infty(\Delta)^{k+m},
\]
where $\mathbb V(\cdot)$ is a mean-zero tight Gaussian process.
In addition,
\[
\sup_{\delta\in\Delta}
\max
\left\{
\left\|\widehat T(\delta)-T(\delta)\right\|_2,
\left\|\widehat C(\delta)-C(\delta)\right\|_F,
\left\|\widehat d(\delta)-d(\delta)\right\|_2
\right\}
=
O_{\Pb}(n^{-1/2}).
\]
\end{assumption}

\begin{assumption}
\label{ass:unif-sl-strong-reg}
The following conditions hold:
\begin{enumerate}[label=(\roman*),leftmargin=2em]
\item The paths
$\beta^*(\delta)$, $\gamma^*_{ac}(\delta)$, $T(\delta)$,
$C(\delta)$, and $d(\delta)$ are continuous on $\Delta$, $h$ is twice
continuously differentiable on a neighborhood of these paths, and the
active set is constant, $J_0(\delta)=J_0$, with $m=|J_0|$.
\item Uniform LICQ, SC, and the second-order condition hold:
\[
\inf_{\delta\in\Delta}
\sigma_{\min}\{C_{ac}(\delta)\}>0,
\qquad
\inf_{\delta\in\Delta}
\min_{j\in J_0}\gamma_j^*(\delta)>0,
\]
with these conditions vacuous when $m=0$, and, for some $\kappa>0$,
\[
v^\top\nabla_\beta^2
h\{\beta^*(\delta),T(\delta)\}v
\geq
\kappa\|v\|_2^2
\]
uniformly over $\delta\in\Delta$ and all
$v$ satisfying $C_{ac}(\delta)v=0$.
\end{enumerate}
\end{assumption}

Assumptions~\ref{ass:unif-sl-process} and
\ref{ass:unif-sl-strong-reg} are the uniform counterparts of the
pointwise component-expansion and KKT-regularity conditions in
Section~\ref{subsec:estimated-linear-constraints}. The former holds,
for example, under uniformly valid asymptotically linear
representations with negligible nuisance remainders. The latter imposes
standard uniform KKT regularity, ensuring that the local solution map
remains well conditioned over $\Delta$. If the active set changes, the
result may instead be applied on compact subintervals that exclude
transition points.

Define
\[
H_{sl}(\delta)
=
\nabla_{\beta}^2h\{\beta^*(\delta),T(\delta)\},
\quad
J_{sl}(\delta)
=
\begin{pmatrix}
H_{sl}(\delta) & C_{ac}(\delta)^\top\\
C_{ac}(\delta) & 0
\end{pmatrix},
\quad
\mathsf A_{sl}(\delta)
=
\begin{pmatrix}I_k&0\end{pmatrix}
J_{sl}(\delta)^{-1}.
\]
Assumption~\ref{ass:unif-sl-strong-reg} implies
\[
\sup_{\delta\in\Delta}
\|J_{sl}(\delta)^{-1}\|<\infty.
\]

The following theorem gives the corresponding uniform first-order
expansion.
\begin{theorem}
\label{thm:unif-linearization-sl}
Suppose Assumptions~\ref{ass:unif-sl-process} and
\ref{ass:unif-sl-strong-reg} hold. Then, with probability tending to
one, for every $\delta\in\Delta$ there exists a unique KKT pair
$(\widehat\beta(\delta),\widehat\gamma_{ac}(\delta))$ in a common
neighborhood of
$(\beta^*(\delta),\gamma^*_{ac}(\delta))$, with active set $J_0$, such
that
\[
\sup_{\delta\in\Delta}
\left\|
\begin{pmatrix}
\widehat\beta(\delta)-\beta^*(\delta)\\
\widehat\gamma_{ac}(\delta)-\gamma^*_{ac}(\delta)
\end{pmatrix}
\right\|_2
=
O_{\Pb}(n^{-1/2}).
\]
Moreover, in $\ell^\infty(\Delta)^{k+m}$,
\[
n^{1/2}
\begin{pmatrix}
\widehat\beta(\delta)-\beta^*(\delta)\\
\widehat\gamma_{ac}(\delta)-\gamma^*_{ac}(\delta)
\end{pmatrix}
=
-
J_{sl}(\delta)^{-1}\mathbb V_n(\delta)
+
o_{\Pb}(1).
\]
Consequently, in $\ell^\infty(\Delta)^k$,
\[
n^{1/2}\{\widehat\beta(\delta)-\beta^*(\delta)\}
=
-
\mathsf A_{sl}(\delta)\mathbb V_n(\delta)
+
o_{\Pb}(1).
\]
Finally, $\widehat\beta(\delta)$ is a strict local minimizer of the
sample program for every $\delta\in\Delta$.
\end{theorem}

\subsection{Bootstrap Simultaneous Bands}
\label{subsec:bands}

Theorems~\ref{thm:unif-linearization} and
\ref{thm:unif-linearization-sl} reduce uniform inference for
$\beta^*(\cdot)$ to inference for suprema of empirical-process
perturbations. We therefore apply standard Gaussian-approximation and
multiplier-bootstrap arguments for empirical-process suprema
\citep[e.g.,][]{chernozhukov2014gaussian,
chernozhukov2016empirical}. Both theorems yield
\[
n^{1/2}\{\widehat\beta(\delta)-\beta^*(\delta)\}
=
\mathbb T_n(\delta)+o_{\Pb}(1)
\quad\text{in}\quad
\ell^\infty(\Delta)^k,
\]
where $\mathbb T_n=-\mathsf A_f\mathbb U_n$ in the fixed-feasible-set
case and $\mathbb T_n=-\mathsf A_{sl}\mathbb V_n$ in the stochastic
linear-constraint case.

Following the standard multiplier-bootstrap construction, let
$\{\xi_i\}_{i=1}^n$ be i.i.d.\ mean-zero, unit-variance multipliers,
independent of the data. In the fixed-feasible-set case, define
\[
\mathbb U_n^\star(\delta)
=
n^{-1/2}
\sum_{i=1}^n
\xi_i
\left[
\widehat G\{Z_i;\delta,\widehat\beta(\delta)\}
-
\Pn\{\widehat G(Z;\delta,\widehat\beta(\delta))\}
\right],
\]
and set
\[
\mathbb T_n^\star(\delta)
=
-\widehat{\mathsf A}_f(\delta)
\mathbb U_n^\star(\delta),
\]
where $\widehat{\mathsf A}_f(\delta)$ is the plug-in estimator of
$\mathsf A_f(\delta)$.
In the stochastic linear-constraint case, define
\[
\widehat{\mathsf A}_{sl}(\delta)
=
\begin{pmatrix}I_k&0\end{pmatrix}
\begin{pmatrix}
\nabla_\beta^2 h\{\widehat\beta(\delta),\widehat T(\delta)\}
&
\widehat C_{ac}(\delta)^\top\\
\widehat C_{ac}(\delta)&0
\end{pmatrix}^{-1}.
\]
Applying the same multiplier construction to the estimated
influence-function representation of
$(\widehat T,\widehat C,\widehat d)$ gives
$\mathbb V_n^\star(\delta)$, and we set
\[
\mathbb T_n^\star(\delta)
=
-\widehat{\mathsf A}_{sl}(\delta)\mathbb V_n^\star(\delta).
\]

Under the usual conditions for Gaussian and multiplier-bootstrap
approximation of empirical-process suprema, together with uniform
consistency of the estimated influence functions and linearization maps,
the conditional distribution of $\mathbb T_n^\star(\cdot)$ consistently
approximates that of
$n^{1/2}\{\widehat\beta(\cdot)-\beta^*(\cdot)\}$;
see \citet[][Theorem~2.1]{chernozhukov2014gaussian} and \citet[][Theorem~2.2]{chernozhukov2016empirical}. Thus, if
$c_{1-\alpha,j}$ is the conditional $(1-\alpha)$ quantile of $\sup_{\delta\in\Delta}
|\mathbb T_{n,j}^\star(\delta)|$, then
\[
\beta_j^*(\delta)
\in
\left[
\widehat\beta_j(\delta)
\pm
n^{-1/2}c_{1-\alpha,j}
\right],
\quad
\delta\in\Delta,
\]
has asymptotic simultaneous coverage $1-\alpha$.

For a fixed covariate value $x$, the uniform delta method gives
\[
n^{1/2}
\left[
f\{x;\widehat\beta(\delta)\}
-
f\{x;\beta^*(\delta)\}
\right]
=
\nabla_\beta f\{x;\beta^*(\delta)\}^\top
\mathbb T_n(\delta)
+
o_{\Pb}(1),
\]
provided $\nabla_\beta f(x;\beta)$ is uniformly continuous and bounded
near the solution path. A simultaneous band for
$\delta\mapsto f\{x;\beta^*(\delta)\}$ is therefore obtained from the
bootstrap process
\[
\mathbb T_{n,x}^\star(\delta)
=
\nabla_\beta f\{x;\widehat\beta(\delta)\}^\top
\mathbb T_n^\star(\delta).
\]

When $1\in\Delta$, simultaneous inference for the contrast
\[
D_x(\delta)
=
f\{x;\beta^*(\delta)\}
-
f\{x;\beta^*(1)\}
\]
follows analogously from
\[
\mathbb T_{n,D}^\star(\delta)
=
\mathbb T_{n,x}^\star(\delta)
-
\mathbb T_{n,x}^\star(1).
\]
In practice, the bands may be evaluated on a fine prespecified grid in
$\Delta$. The theory applies on intervals over which the population
active set is stable.

\section{Numerical Illustration} \label{sec:experiments}
\subsection{Experiments}

\begin{figure}[t!]
\centering
\subfigure[]{\includegraphics[width=0.475\linewidth]{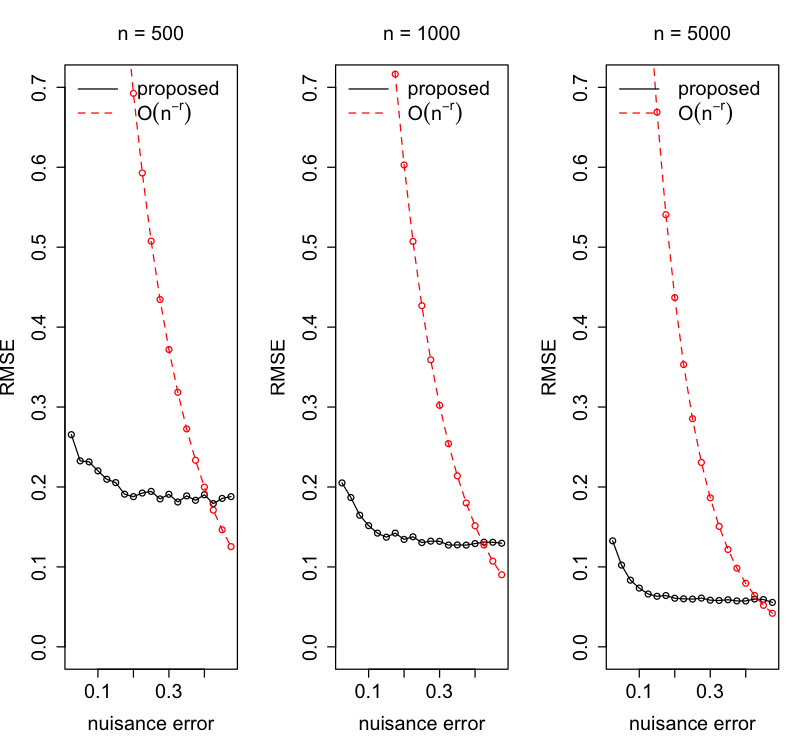}} 
\hfill%
\subfigure[]{\includegraphics[width=0.475\linewidth]{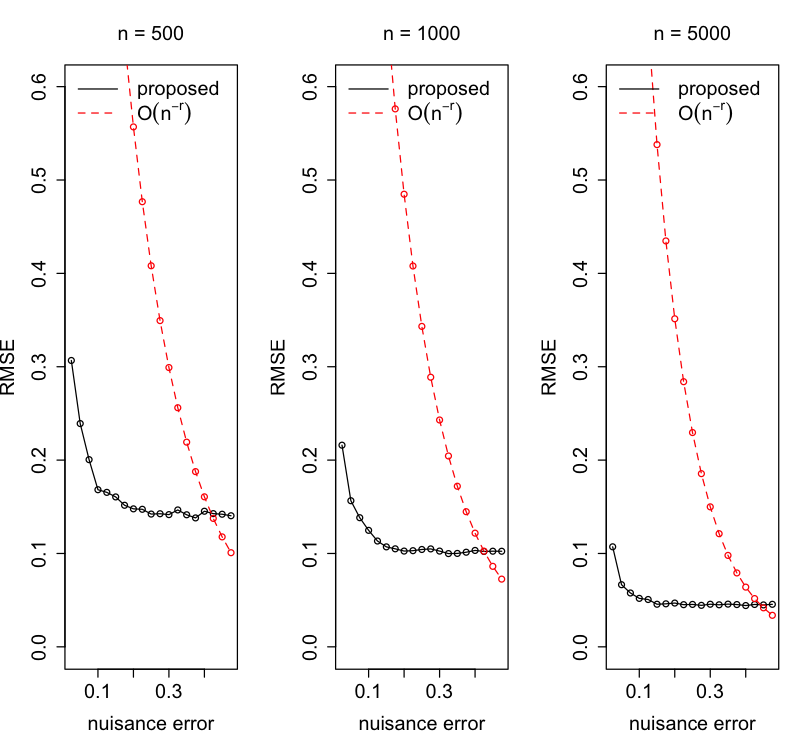}}
\hfill%
\caption{RMSE of the proposed estimator across nuisance convergence
rates for (a) $\delta=0.1$ and (b) $\delta=0.01$. Dashed curves show
the reference nuisance rate $O(n^{-r})$.}
\label{fig:simulation-study}
\end{figure}

\begin{table}[t]
\centering
\setlength{\tabcolsep}{4pt}
\begin{tabular}{cccccccc}
\toprule
& & \multicolumn{3}{c}{$\delta=0.1$} & \multicolumn{3}{c}{$\delta=0.01$} \\
\cmidrule(lr){3-5}\cmidrule(lr){6-8}
$n$ & Estimator & $r=.05$ & $r=.30$ & $r=.50$ & $r=.05$ & $r=.30$ & $r=.50$ \\
\midrule
$500$
& Proposed & $0.27$ & $0.19$ & $0.19$ & $0.30$ & $0.14$  & $0.14$  \\
& IPW      & $0.69$ & $0.25$ & $0.10$ & $0.74$ & $0.37$  & $0.12$  \\
\addlinespace
$1000$
& Proposed & $0.20$ & $0.13$ & $0.13$ & $0.21$ & $0.10$  & $0.10$  \\
& IPW      & $0.69$ & $0.24$ & $0.07$ & $0.74$ & $0.35$  & $0.09$  \\
\addlinespace
$5000$
& Proposed & $0.13$ & $0.06$ & $0.06$ & $0.11$ & $0.045$ & $0.045$ \\
& IPW      & $0.68$ & $0.19$ & $0.03$ & $0.72$ & $0.29$  & $0.041$  \\
\bottomrule
\end{tabular}
\caption{Representative RMSE values for the proposed estimator and an IPW estimator at selected nuisance convergence rates. The IPW estimator uses the estimated incremental intervention weights but does not include the outcome-regression augmentation term, so its error closely tracks the nuisance rate $O(n^{-r})$ in this controlled design.}
\label{tab:simulation-rmse-summary}
\end{table}

We conduct a simulation study to assess the performance of the proposed estimators, emphasizing the validation of the theoretical results established in Section~\ref{sec:analysis}. We focus on the canonical setting described in Example~\ref{example:l2-loss}, using $L_2$ loss and statistical parity as the fairness criterion, with a threshold of $\epsilon=0.1$. The data-generating process is specified as follows: $F \sim \text{Bernoulli}(0.5)$, $X_1 \sim \text{Uniform}[0, 10]$, $X_2 \sim N(4F-2,2)$, $\pi(X) = \text{expit}(2.5-0.3X_1-FX_2)$, and $\mu_A(X)=0.5\sqrt{X_1}+2X_2-5A$ where $X=(X_1,X_2)$. We construct the nuisance estimators as $\widehat{\pi}(X)=\text{expit}\left\{\text{logit}(\pi(X))+\epsilon_\pi \right\}$ and $\widehat{\mu}_a(X) = \mu_a(X)+\epsilon_\mu$, where $\epsilon_\pi, \epsilon_\mu \sim N(n^{-r}, n^{-2r})$. These choices guarantee that $\Vert \widehat{\pi}-\pi\Vert_{2,\Pb}=O(n^{-r})$ and $\Vert \widehat{\mu}_a - \mu_a \Vert_{2,\Pb}=O(n^{-r})$, where we vary $r \in (0,0.5)$ for different scenarios. This simulation setup allows us to study how the proposed semiparametric estimators perform under different convergence rates of the nuisance components. 

For each pair of $(n,\delta)$, where $n\in\{500, 1000, 5000\}$ and $\delta \in \{0.1, 0.01\}$, we generate data and compute nuisance estimates as described above, with $r \in \{0.05 + 0.025k: k=0,\ldots,18\}$. We then compute the coefficients of the proposed estimator $\widehat{\beta}$ by solving the corresponding constrained quadratic program. Each scenario is replicated $500$ times, and the root-mean-square error (RMSE) is computed with respect to the true $\beta^*$. As a simple baseline, we also compute an inverse probability weighting (IPW) estimator that uses the estimated incremental intervention weights but omits the outcome-regression augmentation term. Figure~\ref{fig:simulation-study} and Table~\ref{tab:simulation-rmse-summary} summarize the results. The proposed estimator is substantially less sensitive to nuisance
error than the IPW estimator. For nuisance rates satisfying the
second-order product condition, which in this design corresponds to
$r>1/4$, its RMSE decreases toward the root-$n$ scale as $n$ grows,
whereas the IPW estimator remains more sensitive to propensity
estimation error. These patterns support the role of the efficient influence function correction in reducing nuisance induced bias.

\subsection{Case Study: SMS Reminders and Appointment No-Shows} \label{subsec:case-study}

We apply the proposed method to the {No-Show Appointments} data set,\footnote{Originally compiled by Joni Hoppen (Aquarela Analytics): \url{https://aquare.la/}.} which contains over 110{,}000 scheduled visits in Brazil. Missed appointments create direct revenue loss and idle capacity, and they also generate knock-on costs through rescheduling and inefficient staffing. Clinics therefore often deploy SMS reminders as a low-cost operational lever, yet reminder intensity can shift abruptly due to budget cycles, vendor pricing, staffing constraints, or policy changes. Our objective is to predict no-show risk under such intervention-driven shifts, rather than under the historical reminder policy.

Let $Y\in\{0,1\}$ be the no show indicator and let $A\in\{0,1\}$ indicate whether an SMS reminder was sent. Covariates $X$ include demographics, comorbidities, and appointment characteristics that may confound both reminder assignment and attendance. We consider incremental interventions indexed by $\delta$, which shift the odds of sending SMS reminders relative to current practice. Values $\delta<1$ and $\delta>1$ decrease and increase, respectively, the odds of sending a reminder at every covariate profile. Under the identification assumptions in Section~2.1,
$\E\{Y^{Q(\delta)}\mid X=x\}$ is the counterfactual no-show risk for
patients with characteristics $X=x$ under the prespecified reminder
regime $Q(\delta)$. Thus, $f\{x;\widehat\beta(\delta)\}$ is a counterfactual no show risk score. Ranking patients by this score is meaningful for counterfactual risk stratification, since it identifies patients or subgroups that remain at high risk under the chosen SMS intensity.

\begin{figure}[t!]
\centering
\includegraphics[width=0.80\linewidth]{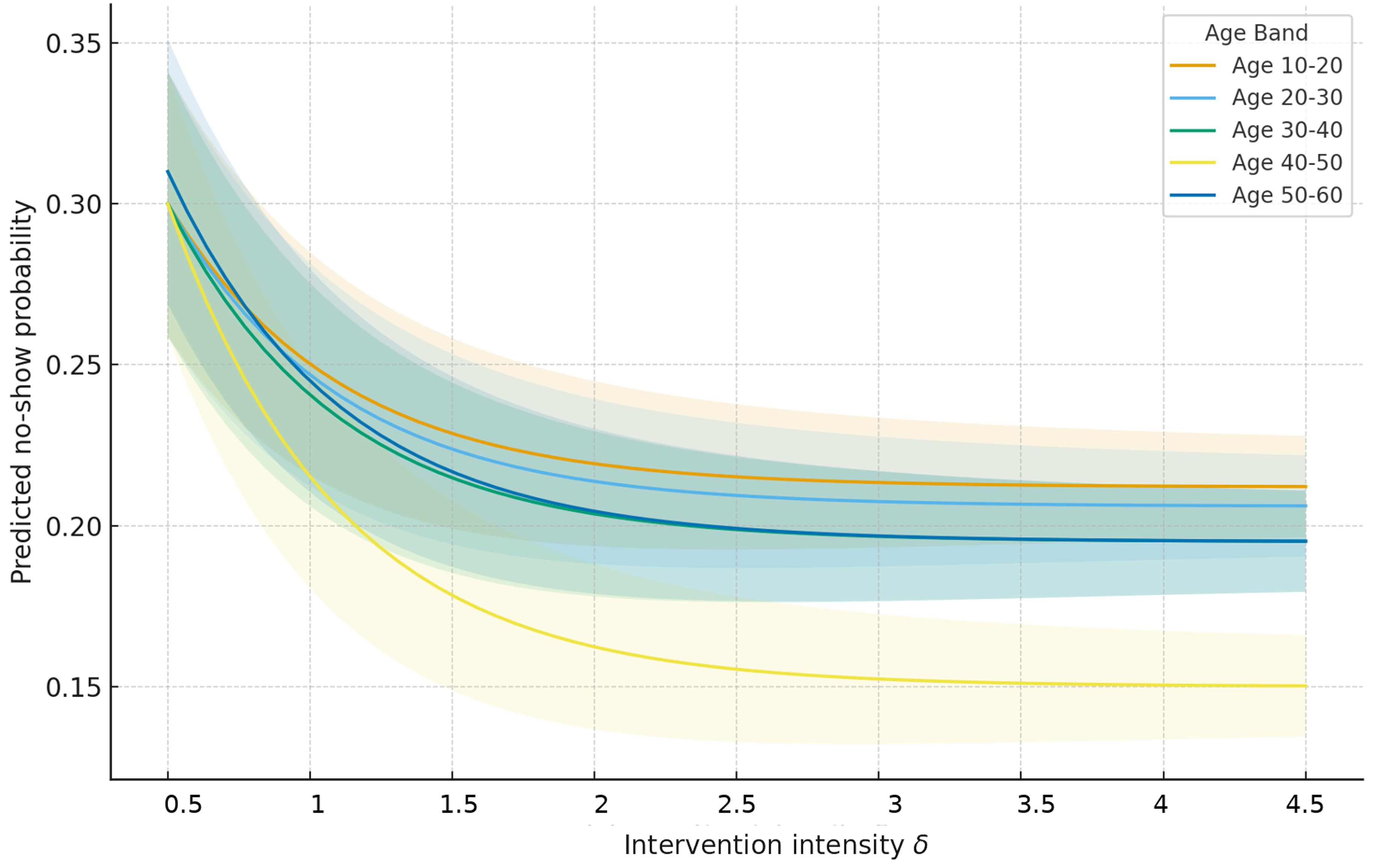}
\caption{Counterfactual prediction curves
$\delta\mapsto f\{x;\widehat\beta(\delta)\}$ with 95\% pointwise
confidence bands. Covariates $x$ are fixed at a representative profile
with only the age band varying, and the vertical axis gives the
predicted counterfactual no-show probability.}
\label{fig:ageband_combined}
\end{figure}

Figure~\ref{fig:ageband_combined} reports
$\delta\mapsto f\{x;\widehat\beta(\delta)\}$ for profiles $x$ that hold
fixed gender (female), scholarship status (none), recorded
comorbidities (none), and waiting time (within one week), while varying
only the age band. We focus on age because it is routinely available at
deployment and provides an interpretable axis of heterogeneity. We
estimate $\widehat\beta(\delta)$ under $L_2$ loss with zero penalty using a linear aggregation, where monotonicity in waiting time is imposed through adjacent-difference constraints on the
relevant coefficients. The nuisance
functions are estimated by random forests with two-fold cross-fitting,
and pointwise confidence intervals are constructed by bootstrap.
Across age bands, higher reminder intensity corresponds to lower fitted
counterfactual no-show risk, with the largest reduction observed for
profiles aged 40--50.

We interpret $f\{x;\widehat\beta(\delta)\}$ as a counterfactual
risk-profiling score under a specified reminder regime, rather than as
an optimal treatment-assignment rule. Comparisons across values of
$\delta$ describe how the fitted risk profile changes across reminder
regimes; see Remark~\ref{rmk:conditional-effects} for direct regime
contrasts.

\section{Discussion}

We developed a framework for counterfactual regression under intervention driven shifts, using incremental interventions as the main regime class while also allowing deterministic interventions as limiting cases. The target is a possibly constrained and penalized risk minimizer, $\beta^*(\delta)$, which induces a counterfactual scoring rule $x\mapsto f\{x;\beta^*(\delta)\}$ under the prespecified regime $Q(\delta)$. Unlike the marginal mean $\E\{Y^{Q(\delta)}\}$, this target provides covariate specific risk profiles and allows losses, penalties, and constraints to encode scientific or operational requirements such as asymmetric costs, shape restrictions, calibration, or fairness. The proposed influence function based estimators provide pointwise and uniform inference for these constrained counterfactual regression targets.

This formulation is complementary to policy learning. The score $f\{x;\beta^*(\delta)\}$ is a regime specific prediction rule, not by itself an optimal treatment allocation rule; when treatment prioritization is the goal, the same machinery can instead be applied to identified contrasts of regime-specific conditional means or to differences between separately estimated counterfactual prediction rules. Incremental interventions can also be viewed as a smooth class of fractional stochastic regimes, connecting the present framework to treatment choice and policy targeting \citep{kitagawa2018should, kitagawa2023treatment}. A natural direction for future work, therefore, is to combine the present inferential framework with policy targeting objectives, for example by optimizing over $\delta$ or over a class of stochastic regimes subject to welfare, resource, regret, or fairness constraints. Other extensions include continuous or time varying treatments, nonsmooth objectives, more complex data adaptive constraints, and integration with conditional effect modeling or optimal treatment regime learning.


\section*{Code Availability}
\texttt{R} scripts for the motivating illustration and simulation study
are available at
\url{https://github.com/kwangho-joshua-kim/counterfactual-prediction}.


\begin{appendices}
\numberwithin{equation}{section}
\section{Additional Technical Results}

\subsection{Regularity Conditions for Assumption \ref{assumption:grad-rootn-CLT-fixed}} \label{appsec:regularity-grad-rootn-CLT-fixed}

Here, we provide high-level sufficient conditions for Assumption \ref{assumption:grad-rootn-CLT-fixed} to hold.

\begin{enumerate}[label=(R\arabic*), leftmargin=2.2em]
\item \textbf{Differentiability and interchange.}
For $\Pb$-a.e.\ $z$, the map $\beta\mapsto \varphi_{\mathcal R}(z;\eta_{\mathcal R},\delta,\beta)$ is differentiable in a neighborhood of $\beta^*$, with derivative
$G(z;\eta_{\mathcal R},\delta,\beta)= \nabla_\beta \varphi_{\mathcal R}(z;\eta_{\mathcal R},\delta,\beta)$.
Moreover, there exist a neighborhood $U$ of $\beta^*$ and an integrable envelope $M(Z)$ with $\E\{M(Z)\}<\infty$ such that
\[
\sup_{\beta\in U}\big\|G(Z;\eta_{\mathcal R},\delta,\beta)\big\|_2 \le M(Z)\quad\text{a.s.},
\]
so that $\nabla_\beta \E\{\varphi_{\mathcal R}(Z;\eta_{\mathcal R},\delta,\beta)\}=\E\{G(Z;\eta_{\mathcal R},\delta,\beta)\}$ for all $\beta\in U$.

\item \textbf{Finite variance.}
$\E\big\|G(Z;\eta_{\mathcal R},\delta,\beta^*)\big\|_2^2<\infty$.

\item \textbf{Penalty regularity.}
$\rho$ is differentiable at $\beta^*$ and $\nabla_\beta\{\lambda\rho(\beta)\}$ is deterministic.

\item \textbf{Negligible nuisance contribution.}
Under cross-fitting, assume
\begin{align}
\left\|
G(Z;\widehat\eta_{\mathcal R},\delta,\beta^*)
-
G(Z;\eta_{\mathcal R},\delta,\beta^*)
\right\|_{2,\Pb}
&=
o_{\Pb}(1),
\\
\left\|
\Pb\left[
G(Z;\widehat\eta_{\mathcal R},\delta,\beta^*)
-
G(Z;\eta_{\mathcal R},\delta,\beta^*)
\right]
\right\|_2
&=
o_{\Pb}(n^{-1/2}).
\end{align}
\end{enumerate}

Note that under cross-fitting,
\[
\Pn(G_{\widehat\eta}-G_\eta)
=
(\Pn-\Pb)(G_{\widehat\eta}-G_\eta)
+
\Pb(G_{\widehat\eta}-G_\eta).
\]
The first term is $o_{\Pb}(n^{-1/2})$ under the first condition in
(R4), while the second is controlled by its second condition. For the incremental-intervention examples considered below, the
second-order bias generally contains both a propensity-error square
and a propensity--outcome-regression product. Let
\[
\nu_a(X)
=
\E\left[
\nabla_\beta L\{Y,f(X;\beta^*)\}
\mid X,A=a
\right],
\quad a\in\mathcal{A},
\]
denote the conditional regression entering the corresponding gradient.
A sufficient rate condition is then
\[
\|\widehat\pi-\pi\|_{2,\Pb}
\left\{
\|\widehat\pi-\pi\|_{2,\Pb}
+
\max_{a\in\{0,1\}}
\|\widehat\nu_a-\nu_a\|_{2,\Pb}
\right\}
=
o_{\Pb}(n^{-1/2});
\]
see \citet{kennedy2019nonparametric}.

\section{Proofs} \label{appsec:proofs}

\textbf{Extra notation.} First, we introduce some extra notation used throughout in the proofs. We let $\langle M_1,M_2\rangle
=
\operatorname{tr}(M_1^\top M_2)$ for conformable matrices $M_1$ and $M_2$, so that
$\|M_1\|_F=\{\langle M_1,M_1\rangle\}^{1/2}$. We let $\mathbb{B}_{\delta}(\bar{z})$ denote the open ball with radius $\delta > 0$ around the point $\bar{z}$ with  $\Vert \cdot \Vert_2$ (unless otherwise mentioned), i.e., $\mathbb{B}_{\delta}(\bar{z})= \{z \mid \Vert z - \bar{z} \Vert_2 < \delta \}$. We use $C^r(\mathbb{S})$ to denote a set of functions that are $r$ times continuously differentiable on $\mathbb{S}$. Further, we let $o_{x \rightarrow \alpha}(1)$ denote some function of $x$ that converges to $0$ when $x \rightarrow \alpha$, i.e., a function $\mathsf{h}_{\alpha}(x)$ such that $\underset{x \rightarrow \alpha}{\lim} \Vert{\mathsf{h}}_{\alpha}(x)\Vert = 0$. This is to distinguish from typical little-o ($o$) asymptotic notation with respect to $n$ in our proofs. Finally, for $\theta \in \R^{p}$ and any vector-valued function $H : \R^{p} \rightarrow \R^{q}$ whose first-order partial derivatives exist, we denote its Jacobian matrix with respect to $\theta$ by $\bm{J}_\theta H \in \R^{q \times p}$. 

\subsection{Proof of Theorem \ref{thm:asymptotics-fixed}}

The proof follows the standard perturbation argument
for smooth constrained stochastic programs
\citep[e.g.,][]{shapiro1993asymptotic,shapiro2000statistical}. Rather than assuming primal--dual localization, we first derive it from the stated regularity conditions before applying this argument.

\begin{proof}
For brevity, write $J_0=J_0(\beta^*)$ and $m=|J_0|$. Let
\[
g_{ac}(\beta)
=
\bigl(g_j(\beta)\bigr)_{j\in J_0}
\in\R^m,
\qquad
\gamma_{ac}
=
\bigl(\gamma_j\bigr)_{j\in J_0}
\in\R^m,
\]
and define
\[
B_{ac}(\beta)
=
\bigl(\nabla_\beta g_j(\beta)^\top\bigr)_{j\in J_0}
\in\R^{m\times k}.
\]

\smallskip

\noindent
By global optimality of $\widehat\beta$ and
Assumption~\ref{assumption:localization-fixed},
\[
0
\leq
\uppsi(\widehat\beta)-\uppsi(\beta^*)
\leq
2\sup_{\beta\in\mathcal C}
\left|
\widehat\uppsi(\beta)-\uppsi(\beta)
\right|
=
o_{\Pb}(1).
\]
The separation condition in
Assumption~\ref{assumption:localization-fixed} therefore implies
\[
\widehat\beta\xrightarrow{p}\beta^*.
\]

Let
\[
\widehat J_0
=
\{j:g_j(\widehat\beta)=0\}
\]
denote the sample active set. For every $j\notin J_0$,
$g_j(\beta^*)<0$, so continuity and primal consistency imply
\[
\Pb(\widehat J_0\subseteq J_0)\longrightarrow1.
\]
Since the gradients indexed by $J_0$ are linearly independent at
$\beta^*$, every subset remains linearly independent in a sufficiently
small neighborhood of $\beta^*$. Hence, with probability tending to
one, LICQ holds at $\widehat\beta$, the sample KKT multiplier vector is
unique, and the KKT conditions hold.

Let $\widehat\gamma_{ac}\in\R^m$ collect the multipliers over $J_0$,
with zero entries assigned to constraints in
$J_0\setminus\widehat J_0$. The sample stationarity equation can then
be written as
\[
\nabla_\beta\widehat\uppsi(\widehat\beta)
+
B_{ac}(\widehat\beta)^\top\widehat\gamma_{ac}
=
0.
\]
Because $B_{ac}(\widehat\beta)$ has full row rank with probability
tending to one,
\[
\widehat\gamma_{ac}
=
-
\left[
B_{ac}(\widehat\beta)B_{ac}(\widehat\beta)^\top
\right]^{-1}
B_{ac}(\widehat\beta)
\nabla_\beta\widehat\uppsi(\widehat\beta).
\]
Moreover,
\[
\begin{aligned}
\nabla_\beta\widehat\uppsi(\widehat\beta)
-
\nabla_\beta\uppsi(\beta^*)
&=
\nabla_\beta\widehat\uppsi(\beta^*)
-
\nabla_\beta\uppsi(\beta^*)
\\
&\quad+
\left[
\int_0^1
\nabla_\beta^2\widehat\uppsi
\{\beta^*+t(\widehat\beta-\beta^*)\}
\,dt
\right]
(\widehat\beta-\beta^*).
\end{aligned}
\]
The first term is $o_{\Pb}(1)$ by
Assumption~\ref{assumption:grad-rootn-CLT-fixed}, while the second is
$o_{\Pb}(1)$ by primal consistency and the local Hessian condition in
Assumption~\ref{assumption:localization-fixed}. Thus,
\[
\nabla_\beta\widehat\uppsi(\widehat\beta)
\xrightarrow{p}
\nabla_\beta\uppsi(\beta^*),
\]
and LICQ yields
\[
\widehat\gamma_{ac}
\xrightarrow{p}
\gamma^*_{ac}.
\]
Finally, SC gives $\gamma_j^*>0$ for every
$j\in J_0$. Hence
$\widehat\gamma_j>0$ for all $j\in J_0$ with probability tending to
one, and complementary slackness implies
\[
\widehat J_0=J_0
\]
with probability tending to one.

Next, define the KKT mapping
$\mathcal K:\R^k\times\R^m\to\R^{k+m}$ by
\[
\mathcal K(\beta,\gamma_{ac})
=
\begin{pmatrix}
\nabla_\beta\uppsi(\beta)
+
B_{ac}(\beta)^\top\gamma_{ac}\\
g_{ac}(\beta)
\end{pmatrix},
\]
so that
$\mathcal K(\beta^*,\gamma_{ac}^*)=0$ at the population solution.
Likewise, define
\[
\widehat{\mathcal K}(\beta,\gamma_{ac})
=
\begin{pmatrix}
\nabla_\beta\widehat\uppsi(\beta)
+
B_{ac}(\beta)^\top\gamma_{ac}\\
g_{ac}(\beta)
\end{pmatrix},
\]
so that the empirical solution satisfies
$\widehat{\mathcal K}(\widehat\beta,\widehat\gamma_{ac})=0$.

Let
\[
\theta
=
\begin{pmatrix}
\beta\\
\gamma_{ac}
\end{pmatrix},
\quad
\theta^*
=
\begin{pmatrix}
\beta^*\\
\gamma^*_{ac}
\end{pmatrix},
\quad
\widehat\theta
=
\begin{pmatrix}
\widehat\beta\\
\widehat\gamma_{ac}
\end{pmatrix}.
\]
With probability tending to one,
$\widehat{\mathcal K}(\widehat\theta)=0$ and
$\mathcal K(\theta^*)=0$. By the integral mean-value expansion,
\[
0
=
\widehat{\mathcal K}(\widehat\theta)-\mathcal K(\theta^*)
=
\widehat{\mathcal K}(\theta^*)-\mathcal K(\theta^*)
+
\overline J_n(\widehat\theta-\theta^*),
\]
where
\[
\overline J_n
=
\int_0^1
D\widehat{\mathcal K}
\{\theta^*+t(\widehat\theta-\theta^*)\}
\,dt.
\]
By primal--dual consistency established above, the local Hessian
condition in Assumption~\ref{assumption:localization-fixed}, and
continuity of the constraint derivatives,
\[
\overline J_n
\xrightarrow{p}
J_f(\beta^*)
=
\begin{pmatrix}
H(\beta^*) & B_{ac}^\top\\
B_{ac} & 0
\end{pmatrix}.
\]
Assumptions~\ref{assumption:LICQ-SC} and
\ref{assumption:quadratic-growth} imply that
$J_f(\beta^*)$ is nonsingular. Moreover,
\[
\widehat{\mathcal K}(\theta^*)-\mathcal K(\theta^*)
=
\begin{pmatrix}
\nabla_\beta\widehat\uppsi(\beta^*)
-
\nabla_\beta\uppsi(\beta^*)
\\
0
\end{pmatrix}.
\]
Therefore,
\[
n^{1/2}(\widehat\theta-\theta^*)
=
-
\overline J_n^{-1}
\begin{pmatrix}
n^{1/2}
\{
\nabla_\beta\widehat\uppsi(\beta^*)
-
\nabla_\beta\uppsi(\beta^*)
\}
\\
0
\end{pmatrix}.
\]
Assumption~\ref{assumption:grad-rootn-CLT-fixed} and Slutsky's theorem
give
\[
n^{1/2}
\begin{pmatrix}
\widehat\beta-\beta^*\\
\widehat\gamma_{ac}-\gamma^*_{ac}
\end{pmatrix}
\xrightarrow{d}
-
J_f(\beta^*)^{-1}
\begin{pmatrix}
W\\
0
\end{pmatrix}.
\]
Taking the first $k$ coordinates yields the stated marginal limit.
\end{proof}

\subsection{Proof of Theorem~\ref{thm:rates}}
\label{proof:rates}

\begin{proof}
For parameter values $(T',C',d')$ near $(T,C,d)$, the linear
constraints allow the estimated program to be viewed as a
finite-dimensional perturbation of the population program. Its
active-set KKT map is
\[
F_{sl}(\beta,\gamma_{ac};T',C',d')
=
\begin{pmatrix}
\nabla_\beta h(\beta,T')
+
C'_{ac}{}^\top\gamma_{ac}
\\
C'_{ac}\beta-d'_{ac}
\end{pmatrix}.
\]
At the population KKT pair,
\[
F_{sl}(\beta^*,\gamma^*_{ac};T,C,d)=0.
\]
For nearby $(T',C',d')$, the perturbation of the KKT equations at the
population solution is exactly
\[
F_{sl}(\beta^*,\gamma^*_{ac};T',C',d')
-
F_{sl}(\beta^*,\gamma^*_{ac};T,C,d)
=
\begin{pmatrix}
\nabla_\beta h(\beta^*,T')
-
\nabla_\beta h(\beta^*,T)
+
(C'_{ac}-C_{ac})^\top\gamma^*_{ac}
\\[2mm]
(C'_{ac}-C_{ac})\beta^*
-
(d'_{ac}-d_{ac})
\end{pmatrix}.
\]
Thus, the constraint perturbation enters through the finite-dimensional
coefficients $(C',d')$ without a nonlinear remainder. In particular,
\[
D_\beta(C'_{ac}\beta-d'_{ac})=C'_{ac},
\qquad
D_\beta^2(C'_{ac}\beta-d'_{ac})=0.
\]
The Jacobian with respect to $(\beta,\gamma_{ac})$ is therefore
\[
D_{(\beta,\gamma_{ac})}
F_{sl}(\beta^*,\gamma^*_{ac};T,C,d)
=
M
=
\begin{pmatrix}
\nabla_\beta^2h(\beta^*,T) & C_{ac}^\top\\
C_{ac} & 0
\end{pmatrix}.
\]
Assumption~\ref{assumption:strong-reg-sl} implies that $M$ is
nonsingular. The implicit-function theorem therefore yields
neighborhoods of $(T,C,d)$ and $(\beta^*,\gamma^*_{ac})$ and a unique
continuously differentiable solution map
\[
(T',C',d')
\longmapsto
\{\beta(T',C',d'),\gamma_{ac}(T',C',d')\}.
\]
Because inactive population constraints are strictly slack and
$\gamma^*_{ac}>0$ componentwise, continuity ensures that the inactive
constraints remain slack and the active multipliers remain positive
under sufficiently small perturbations. The resulting solution
therefore satisfies the KKT conditions of the full inequality-constrained
program. The second-order condition in
Assumption~\ref{assumption:strong-reg-sl} also persists locally, so this
solution is a strict local minimizer.

The local solution map is locally Lipschitz. Hence, on an event whose
probability tends to one,
\[
\left\|
\begin{pmatrix}
\widehat\beta-\beta^*\\
\widehat\gamma_{ac}-\gamma^*_{ac}
\end{pmatrix}
\right\|_2
\leq
L
\left(
\|\widehat T-T\|_2
+
\|\widehat C-C\|_F
+
\|\widehat d-d\|_2
\right)
\]
for some finite constant $L$. Since the parameter space is
finite-dimensional, this yields the stated rate under
Assumption~\ref{assumption:consistency}.

If $h(\cdot,T')$ is strongly convex on $\R^k$ for every $T'$ near
$T$, then the sample program is convex and its locally identified KKT
solution is its unique global optimizer.
\end{proof}

\subsection{Proof of Theorem~\ref{thm:asymptotics}}
\label{proof:asymptotics}

\begin{proof}
Let
\[
\theta
=
\begin{pmatrix}
\beta\\
\gamma_{ac}
\end{pmatrix},
\quad
\theta^*
=
\begin{pmatrix}
\beta^*\\
\gamma^*_{ac}
\end{pmatrix},
\quad
\widehat\theta
=
\begin{pmatrix}
\widehat\beta\\
\widehat\gamma_{ac}
\end{pmatrix}.
\]
Define the population and sample active-set KKT maps by
\[
F_{sl}(\theta)
=
\begin{pmatrix}
\nabla_\beta h(\beta,T)+C_{ac}^\top\gamma_{ac}\\
C_{ac}\beta-d_{ac}
\end{pmatrix},
\quad
\widehat F_{sl}(\theta)
=
\begin{pmatrix}
\nabla_\beta h(\beta,\widehat T)
+
\widehat C_{ac}^\top\gamma_{ac}\\
\widehat C_{ac}\beta-\widehat d_{ac}
\end{pmatrix}.
\]
By the population and sample KKT conditions,
\[
F_{sl}(\theta^*)=0,
\quad
\widehat F_{sl}(\widehat\theta)=0.
\]
Moreover, by direct calculation,
\[
n^{1/2}
\{
\widehat F_{sl}(\theta^*)-F_{sl}(\theta^*)
\}
=
\Xi_n.
\]

By Theorem~\ref{thm:rates} and
Assumption~\ref{assumption:root-n-rates},
\[
\|\widehat\theta-\theta^*\|_2
=
O_{\Pb}(n^{-1/2}).
\]

We now make explicit how the linear constraints enter the first-order
expansion. For the stationarity equation,
\begin{align*}
0
&=
\nabla_\beta h(\widehat\beta,\widehat T)
+
\widehat C_{ac}^\top\widehat\gamma_{ac}
-
\left\{
\nabla_\beta h(\beta^*,T)
+
C_{ac}^\top\gamma^*_{ac}
\right\}
\\
&=
\nabla_\beta^2h(\beta^*,T)
(\widehat\beta-\beta^*)
+
C_{ac}^\top
(\widehat\gamma_{ac}-\gamma^*_{ac})
\\
&\quad+
\left[
\nabla_\beta h(\beta^*,\widehat T)
-
\nabla_\beta h(\beta^*,T)
+
(\widehat C_{ac}-C_{ac})^\top\gamma^*_{ac}
\right]
+
o_{\Pb}(n^{-1/2}).
\end{align*}
The cross-product term in the remainder satisfies
\[
\left\|
(\widehat C_{ac}-C_{ac})^\top
(\widehat\gamma_{ac}-\gamma^*_{ac})
\right\|_2
\leq
\|\widehat C_{ac}-C_{ac}\|_F
\|\widehat\gamma_{ac}-\gamma^*_{ac}\|_2
=
O_{\Pb}(n^{-1}),
\]
while the remaining Taylor remainder is
$o_{\Pb}(n^{-1/2})$ by the smoothness of $h$ and the root-$n$
localization established above.

For the active constraints, linearity gives the exact decomposition
\begin{align*}
0
&=
\widehat C_{ac}\widehat\beta-\widehat d_{ac}
-
(C_{ac}\beta^*-d_{ac})
\\
&=
C_{ac}(\widehat\beta-\beta^*)
+
(\widehat C_{ac}-C_{ac})\beta^*
-
(\widehat d_{ac}-d_{ac})
\\
&\quad+
(\widehat C_{ac}-C_{ac})
(\widehat\beta-\beta^*).
\end{align*}
The last term is $O_{\Pb}(n^{-1})$ under
Assumption~\ref{assumption:root-n-rates} and
Theorem~\ref{thm:rates}. Combining the two equations yields
\[
0
=
M
\begin{pmatrix}
\widehat\beta-\beta^*\\
\widehat\gamma_{ac}-\gamma^*_{ac}
\end{pmatrix}
+
n^{-1/2}\Xi_n
+
o_{\Pb}(n^{-1/2}).
\]
Since $M$ is nonsingular,
\[
n^{1/2}
\begin{pmatrix}
\widehat\beta-\beta^*\\
\widehat\gamma_{ac}-\gamma^*_{ac}
\end{pmatrix}
=
-M^{-1}\Xi_n+o_{\Pb}(1).
\]
Taking the first \(k\) coordinates gives
\[
n^{1/2}(\widehat\beta-\beta^*)
=
-
\begin{pmatrix}I_k&0\end{pmatrix}
M^{-1}\Xi_n
+
o_{\Pb}(1).
\]
The weak limit follows from
Assumption~\ref{assumption:kkt-convergence} and Slutsky's theorem.
\end{proof}

\subsection{Proof of Theorem~\ref{thm:unif-linearization}}
\label{app:unif-delta-proofs-fixed}

\begin{proof}
Let $J_0$ be the common active set over $\Delta$ and let
$m=|J_0|$. Define
\[
g_{ac}(\beta)
=
\left(g_j(\beta)\right)_{j\in J_0},
\quad
B_{ac}(\beta)
=
\left(
\nabla_\beta g_j(\beta)^\top
\right)_{j\in J_0}.
\]
For
\[
\theta
=
\begin{pmatrix}
\beta\\
\gamma_{ac}
\end{pmatrix},
\quad
\theta^*(\delta)
=
\begin{pmatrix}
\beta^*(\delta)\\
\gamma^*_{ac}(\delta)
\end{pmatrix},
\]
define the population and empirical active-set KKT maps
\[
F_f(\delta,\theta)
=
\begin{pmatrix}
\nabla_\beta\uppsi_\delta(\beta)
+
B_{ac}(\beta)^\top\gamma_{ac}
\\
g_{ac}(\beta)
\end{pmatrix},
\]
and
\[
\widehat F_f(\delta,\theta)
=
\begin{pmatrix}
\nabla_\beta\widehat\uppsi_\delta(\beta)
+
B_{ac}(\beta)^\top\gamma_{ac}
\\
g_{ac}(\beta)
\end{pmatrix}.
\]
By the population KKT conditions,
\[
F_f\{\delta,\theta^*(\delta)\}=0
\quad
\text{for all }\delta\in\Delta.
\]

At the population solution path,
\[
\widehat F_f\{\delta,\theta^*(\delta)\}
-
F_f\{\delta,\theta^*(\delta)\}
=
\begin{pmatrix}
\nabla_\beta\widehat\uppsi_\delta\{\beta^*(\delta)\}
-
\nabla_\beta\uppsi_\delta\{\beta^*(\delta)\}
\\
0
\end{pmatrix}.
\]
Assumption~\ref{ass:unif-grad-fclt} therefore implies
\[
\sup_{\delta\in\Delta}
\left\|
\widehat F_f\{\delta,\theta^*(\delta)\}
\right\|_2
=
O_{\Pb}(n^{-1/2}).
\]

By Assumption~\ref{ass:unif-strong-reg}, the population KKT Jacobian
\[
D_\theta F_f\{\delta,\theta^*(\delta)\}
=
J_f(\delta)
\]
is uniformly nonsingular. The common-neighborhood smoothness and
uniform Hessian convergence further imply that the empirical KKT
Jacobians are uniformly close to the population KKT Jacobians on a
fixed neighborhood of the solution path. Hence, by the uniform
implicit-function argument, with probability tending to one there
exists, for every $\delta\in\Delta$, a unique zero
\[
\widehat\theta(\delta)
=
\begin{pmatrix}
\widehat\beta(\delta)\\
\widehat\gamma_{ac}(\delta)
\end{pmatrix}
\]
of the empirical active-set KKT map in this neighborhood, satisfying
\[
\sup_{\delta\in\Delta}
\|\widehat\theta(\delta)-\theta^*(\delta)\|_2
=
O_{\Pb}(n^{-1/2}).
\]

Because $\Delta$ is compact, the solution path is continuous, and the
active set is constant, the inactive population constraints are
uniformly slack. Uniform SC and the preceding
localization therefore imply that, with probability tending to one,
the active multipliers remain positive and the inactive constraints
remain slack uniformly over $\Delta$. Thus the empirical and population
active sets agree. The uniform second-order condition in
Assumption~\ref{ass:unif-strong-reg} also persists under the empirical
perturbation, so $\widehat\beta(\delta)$ is a strict local optimizer.

We now derive the first-order expansion. By the integral mean-value
formula,
\[
0
=
\widehat F_f\{\delta,\widehat\theta(\delta)\}
-
F_f\{\delta,\theta^*(\delta)\}
=
\widehat F_f\{\delta,\theta^*(\delta)\}
-
F_f\{\delta,\theta^*(\delta)\}
+
\overline J_{n,f}(\delta)
\{\widehat\theta(\delta)-\theta^*(\delta)\},
\]
where
\[
\overline J_{n,f}(\delta)
=
\int_0^1
D_\theta\widehat F_f
\left[
\delta,\,
\theta^*(\delta)
+
t\{\widehat\theta(\delta)-\theta^*(\delta)\}
\right]
dt.
\]
By Assumption~\ref{ass:unif-strong-reg} and the uniform
$O_{\Pb}(n^{-1/2})$ localization established above,
\[
\sup_{\delta\in\Delta}
\left\|
\overline J_{n,f}(\delta)-J_f(\delta)
\right\|
=
o_{\Pb}(1).
\]
Hence
\[
\sup_{\delta\in\Delta}
\left\|
\overline J_{n,f}(\delta)^{-1}
-
J_f(\delta)^{-1}
\right\|
=
o_{\Pb}(1).
\]

Multiplying the mean-value expansion by $n^{1/2}$ gives
\[
n^{1/2}
\{\widehat\theta(\delta)-\theta^*(\delta)\}
=
-
\overline J_{n,f}(\delta)^{-1}
\begin{pmatrix}
n^{1/2}
\left[
\nabla_\beta\widehat\uppsi_\delta\{\beta^*(\delta)\}
-
\nabla_\beta\uppsi_\delta\{\beta^*(\delta)\}
\right]
\\
0
\end{pmatrix}.
\]
Assumption~\ref{ass:unif-grad-fclt} and the preceding uniform inverse
convergence therefore yield
\[
n^{1/2}
\{\widehat\theta(\delta)-\theta^*(\delta)\}
=
-
J_f(\delta)^{-1}
\begin{pmatrix}
\mathbb U_n(\delta)\\
0
\end{pmatrix}
+
o_{\Pb}(1)
\]
in $\ell^\infty(\Delta)^{k+m}$. Taking the first $k$ coordinates gives
\[
n^{1/2}
\{\widehat\beta(\delta)-\beta^*(\delta)\}
=
-
\begin{pmatrix}
I_k&0
\end{pmatrix}
J_f(\delta)^{-1}
\begin{pmatrix}
I_k\\
0
\end{pmatrix}
\mathbb U_n(\delta)
+
o_{\Pb}(1),
\]
namely,
\[
n^{1/2}
\{\widehat\beta(\delta)-\beta^*(\delta)\}
=
-
\mathsf A_f(\delta)\mathbb U_n(\delta)
+
o_{\Pb}(1).
\]
\end{proof}

\subsection{Proof of Theorem~\ref{thm:unif-linearization-sl}}
\label{app:unif-delta-proofs-sl}

\begin{proof}
Write
\[
\theta^*(\delta)
=
\begin{pmatrix}
\beta^*(\delta)\\
\gamma^*_{ac}(\delta)
\end{pmatrix},
\quad
\widehat\theta(\delta)
=
\begin{pmatrix}
\widehat\beta(\delta)\\
\widehat\gamma_{ac}(\delta)
\end{pmatrix},
\]
and define the reduced KKT map
\[
F_{sl}(\delta,\beta,\gamma;T,C,d)
=
\begin{pmatrix}
\nabla_\beta h\{\beta,T(\delta)\}
+
C_{ac}(\delta)^\top\gamma\\
C_{ac}(\delta)\beta-d_{ac}(\delta)
\end{pmatrix},
\]
with $\widehat F_{sl}$ defined analogously using
$(\widehat T,\widehat C,\widehat d)$. By the population KKT conditions,
\[
F_{sl}\{\delta,\theta^*(\delta);T,C,d\}=0
\]
for every $\delta\in\Delta$.

The Jacobian of $F_{sl}$ with respect to $(\beta,\gamma)$ at
$\theta^*(\delta)$ is $J_{sl}(\delta)$. By
Assumption~\ref{ass:unif-sl-strong-reg},
\[
\sup_{\delta\in\Delta}
\|J_{sl}(\delta)^{-1}\|<\infty.
\]
Assumption~\ref{ass:unif-sl-process} yields a uniformly
$O_{\Pb}(n^{-1/2})$ sample KKT residual at the population path and
uniform consistency of the estimated program components. Together with
the smoothness conditions in
Assumption~\ref{ass:unif-sl-strong-reg}, this also yields uniform
convergence of the sample KKT Jacobian to its population counterpart on
a common neighborhood of the solution path. A uniform implicit-function
argument therefore gives a unique zero $\widehat\theta(\delta)$ of the
reduced sample KKT system in this neighborhood, with
\[
\sup_{\delta\in\Delta}
\|\widehat\theta(\delta)-\theta^*(\delta)\|_2
=
O_{\Pb}(n^{-1/2}).
\]

Because the active set is constant and the population paths are
continuous on the compact set $\Delta$, all inactive constraints are
uniformly slack. Uniform convergence of $(\widehat C,\widehat d)$
therefore preserves their inactivity. Uniform SC and the preceding
localization preserve positivity of the active multipliers, while
uniform LICQ persists under the convergence of $\widehat C_{ac}$ to
$C_{ac}$. Hence the reduced solution is a KKT pair for the full sample
program with active set $J_0$, uniformly over $\delta\in\Delta$.

For the first-order expansion, an exact mean-value expansion of the
sample KKT system gives
\[
0
=
\widehat F_{sl}\{\delta,\theta^*(\delta)\}
+
\overline J_{n,sl}(\delta)
\{\widehat\theta(\delta)-\theta^*(\delta)\},
\]
where $\overline J_{n,sl}(\delta)$ is the sample KKT Jacobian averaged
along the line segment joining $\theta^*(\delta)$ and
$\widehat\theta(\delta)$. The localization above and the uniform
smoothness and component-convergence conditions imply
\[
\sup_{\delta\in\Delta}
\|\overline J_{n,sl}(\delta)-J_{sl}(\delta)\|
=
o_{\Pb}(1).
\]
By Assumption~\ref{ass:unif-sl-process},
\[
n^{1/2}
\widehat F_{sl}\{\delta,\theta^*(\delta)\}
=
\mathbb V_n(\delta)+o_{\Pb}(1)
\]
in $\ell^\infty(\Delta)^{k+m}$. Uniform invertibility and Slutsky's
theorem therefore yield
\[
n^{1/2}
\{\widehat\theta(\delta)-\theta^*(\delta)\}
=
-
J_{sl}(\delta)^{-1}\mathbb V_n(\delta)
+
o_{\Pb}(1)
\]
in $\ell^\infty(\Delta)^{k+m}$. Premultiplying by
$\begin{pmatrix}I_k&0\end{pmatrix}$ gives the stated expansion for
$\widehat\beta(\delta)$.

Finally, uniform convergence of the Hessian and of
$\widehat C_{ac}$, together with the second-order condition in
Assumption~\ref{ass:unif-sl-strong-reg}, implies that, with probability
tending to one,
\[
v^\top
\nabla_\beta^2
h\{\widehat\beta(\delta),\widehat T(\delta)\}v
>0
\]
uniformly over $\delta\in\Delta$ and nonzero $v$ satisfying
$\widehat C_{ac}(\delta)v=0$. Thus the sample second-order sufficient
condition holds at the KKT pair, and
$\widehat\beta(\delta)$ is a strict local minimizer of the sample
program for every $\delta\in\Delta$.
\end{proof}
\end{appendices}


\bibliographystyle{plainnat}
\bibliography{bibliography}

@article{kitagawa2018should,
  title={Who should be treated? empirical welfare maximization methods for treatment choice},
  author={Kitagawa, Toru and Tetenov, Aleksey},
  journal={Econometrica},
  volume={86},
  number={2},
  pages={591--616},
  year={2018},
  publisher={Wiley Online Library}
}

@article{kennedy2020optimal,
  title={Optimal doubly robust estimation of heterogeneous causal effects},
  author={Kennedy, Edward H},
  journal={arXiv preprint arXiv:2004.14497},
  year={2020}
}

@article{kennedy2021semiparametric,
  title={Semiparametric counterfactual density estimation},
  author={Kennedy, Edward H and Balakrishnan, Sivaraman and Wasserman, Larry},
  journal={arXiv preprint arXiv:2102.12034},
  year={2021}
}

@article{shapiro1993asymptotic,
  title={Asymptotic behavior of optimal solutions in stochastic programming},
  author={Shapiro, Alexander},
  journal={Mathematics of Operations Research},
  volume={18},
  number={4},
  pages={829--845},
  year={1993},
  publisher={INFORMS}
}

@book{imbens2015causal,
  title={Causal inference in statistics, social, and biomedical sciences},
  author={Imbens, Guido W and Rubin, Donald B},
  year={2015},
  publisher={Cambridge University Press}
}

@article{Chernozhukov18,
    author = {Chernozhukov, Victor and Chetverikov, Denis and Demirer, Mert and Duflo, Esther and Hansen, Christian and Newey, Whitney and Robins, James},
    title = "{Double/debiased machine learning for treatment and structural parameters}",
    journal = {The Econometrics Journal},
    volume = {21},
    number = {1},
    pages = {C1-C68},
    year = {2018},
    month = {01}
}

@article{schulam2017reliable,
  title={Reliable decision support using counterfactual models},
  author={Schulam, Peter and Saria, Suchi},
  journal={Advances in Neural Information Processing Systems},
  volume={30},
  pages={1697--1708},
  year={2017}
}

@article{nguyen2020counterfactual,
  title={Counterfactual clinical prediction models could help to infer individualized treatment effects in randomized controlled trials—An illustration with the International Stroke Trial},
  author={Nguyen, Tri-Long and Collins, Gary S and Landais, Paul and Le Manach, Yannick},
  journal={Journal of clinical epidemiology},
  volume={125},
  pages={47--56},
  year={2020},
  publisher={Elsevier}
}

@incollection{kennedy2016semiparametric,
  title={Semiparametric theory and empirical processes in causal inference},
  author={Kennedy, Edward H},
  booktitle={Statistical causal inferences and their applications in public health research},
  pages={141--167},
  year={2016},
  publisher={Springer}
}

@article{li2016predictive,
  title={A predictive enrichment procedure to identify potential responders to a new therapy for randomized, comparative controlled clinical studies},
  author={Li, Junlong and Zhao, Lihui and Tian, Lu and Cai, Tianxi and Claggett, Brian and Callegaro, Andrea and Dizier, Benjamin and Spiessens, Bart and Ulloa-Montoya, Fernando and Wei, Lee-Jen},
  journal={Biometrics},
  volume={72},
  number={3},
  pages={877--887},
  year={2016},
  publisher={Wiley Online Library}
}

@book{van2000asymptotic,
  author    = {van der Vaart, A. W.},
  title     = {Asymptotic Statistics},
  publisher = {Cambridge University Press},
  year      = {1998}
}

@article{still2018lectures,
  title={Lectures on parametric optimization: An introduction},
  author={Still, Georg},
  journal={Optimization Online},
  year={2018}
}

@article{dupacova1988asymptotic,
  title={Asymptotic behavior of statistical estimators and of optimal solutions of stochastic optimization problems},
  author={Dupacov{\'a}, Jitka and Wets, Roger},
  journal={The annals of statistics},
  volume={16},
  number={4},
  pages={1517--1549},
  year={1988},
  publisher={Institute of Mathematical Statistics}
}

@article{james2013penalized,
  title={Penalized and constrained regression},
  author={James, Gareth M and Paulson, Courtney and Rusmevichientong, Paat},
  journal={Unpublished manuscript, http://www-bcf. usc. edu/gareth/research/Research. html},
  year={2013},
  publisher={Citeseer}
}

@article{gaines2018algorithms,
  title={Algorithms for fitting the constrained lasso},
  author={Gaines, Brian R and Kim, Juhyun and Zhou, Hua},
  journal={Journal of Computational and Graphical Statistics},
  volume={27},
  number={4},
  pages={861--871},
  year={2018},
  publisher={Taylor \& Francis}
}

@article{lin2021scoping,
  title={A scoping review of causal methods enabling predictions under hypothetical interventions},
  author={Lin, Lijing and Sperrin, Matthew and Jenkins, David A and Martin, Glen P and Peek, Niels},
  journal={Diagnostic and prognostic research},
  volume={5},
  number={1},
  pages={1--16},
  year={2021},
  publisher={Springer}
}

@article{dickerman2020counterfactual,
  title={Counterfactual prediction is not only for causal inference},
  author={Dickerman, Barbra A and Hern{\'a}n, Miguel A},
  journal={European Journal of Epidemiology},
  volume={35},
  number={7},
  pages={615--617},
  year={2020},
  publisher={Springer}
}

@incollection{shapiro2000statistical,
  title={Statistical inference of stochastic optimization problems},
  author={Shapiro, Alexander},
  booktitle={Probabilistic constrained optimization},
  pages={282--307},
  year={2000},
  publisher={Springer}
}

@article{nemirovski2009robust,
  title={Robust stochastic approximation approach to stochastic programming},
  author={Nemirovski, Arkadi and Juditsky, Anatoli and Lan, Guanghui and Shapiro, Alexander},
  journal={SIAM Journal on optimization},
  volume={19},
  number={4},
  pages={1574--1609},
  year={2009},
  publisher={SIAM}
}

@article{mishler2021fade,
      title={FADE: FAir Double Ensemble Learning for Observable and Counterfactual Outcomes}, 
      author={Alan Mishler and Edward Kennedy},
      year={2021},
      eprint={2109.00173},
      archivePrefix={arXiv},
      journal={arXiv preprint arXiv:2109.00173},
      primaryClass={stat.ML}
}

@inproceedings{coston2020RuntimeConfounding,
	title = {Counterfactual Predictions under Runtime Confounding},
	eventtitle = {NeurIPS},
	volume = {33},
	pages = {4150--4162},
	booktitle = {Advances in neural information processing systems},
	author = {Coston, Amanda and Kennedy, Edward and Chouldechova, Alexandra},
	year = {2020},
	langid = {english}
}

@article{lu2019generalized,
  title={Generalized linear models with linear constraints for microbiome compositional data},
  author={Lu, Jiarui and Shi, Pixu and Li, Hongzhe},
  journal={Biometrics},
  volume={75},
  number={1},
  pages={235--244},
  year={2019},
  publisher={Wiley Online Library}
}

@article{wachsmuth2013licq,
  title={On LICQ and the uniqueness of Lagrange multipliers},
  author={Wachsmuth, Gerd},
  journal={Operations Research Letters},
  volume={41},
  number={1},
  pages={78--80},
  year={2013},
  publisher={Elsevier}
}

@article{kennedy2024semiparametric,
  title={Semiparametric doubly robust targeted double machine learning: a review},
  author={Kennedy, Edward H},
  journal={Handbook of Statistical Methods for Precision Medicine},
  pages={207--236},
  year={2024},
  publisher={Chapman and Hall/CRC}
}

@article{kim2022counterfactual,
  title={Counterfactual Mean-variance Optimization},
  author={Kim, Kwangho and Mishler, Alan and Zubizarreta, Jos{\'e} R},
  journal={arXiv preprint arXiv:2209.09538},
  year={2022}
}

@article{van2020prediction,
  title={Prediction meets causal inference: the role of treatment in clinical prediction models},
  author={van Geloven, Nan and Swanson, Sonja A and Ramspek, Chava L and Luijken, Kim and van Diepen, Merel and Morris, Tim P and Groenwold, Rolf HH and van Houwelingen, Hans C and Putter, Hein and le Cessie, Saskia},
  journal={European journal of epidemiology},
  volume={35},
  number={7},
  pages={619--630},
  year={2020},
  publisher={Springer}
}

@article{dickerman2022predicting,
  title={Predicting counterfactual risks under hypothetical treatment strategies: an application to HIV},
  author={Dickerman, Barbra A and Dahabreh, Issa J and Cantos, Krystal V and Logan, Roger W and Lodi, Sara and Rentsch, Christopher T and Justice, Amy C and Hern{\'a}n, Miguel A},
  journal={European journal of epidemiology},
  volume={37},
  number={4},
  pages={367--376},
  year={2022},
  publisher={Springer}
}

@article{sperrin2018using,
  title={Using marginal structural models to adjust for treatment drop-in when developing clinical prediction models},
  author={Sperrin, Matthew and Martin, Glen P and Pate, Alexander and Van Staa, Tjeerd and Peek, Niels and Buchan, Iain},
  journal={Statistics in medicine},
  volume={37},
  number={28},
  pages={4142--4154},
  year={2018},
  publisher={Wiley Online Library}
}

@inproceedings{shalit2017estimating,
  title={Estimating individual treatment effect: generalization bounds and algorithms},
  author={Shalit, Uri and Johansson, Fredrik D and Sontag, David},
  booktitle={International conference on machine learning},
  pages={3076--3085},
  year={2017},
  organization={PMLR}
}

@inproceedings{johansson2016learning,
  title={Learning representations for counterfactual inference},
  author={Johansson, Fredrik and Shalit, Uri and Sontag, David},
  booktitle={International conference on machine learning},
  pages={3020--3029},
  year={2016},
  organization={PMLR}
}

@inproceedings{hassanpour2019counterfactual,
  title={CounterFactual Regression with Importance Sampling Weights.},
  author={Hassanpour, Negar and Greiner, Russell},
  booktitle={IJCAI},
  pages={5880--5887},
  year={2019},
  organization={Macao}
}

@inproceedings{hassanpour2019learning,
  title={Learning disentangled representations for counterfactual regression},
  author={Hassanpour, Negar and Greiner, Russell},
  booktitle={International Conference on Learning Representations},
  year={2019}
}

@article{kennedy2019nonparametric,
  title={Nonparametric causal effects based on incremental propensity score interventions},
  author={Kennedy, Edward H},
  journal={Journal of the American Statistical Association},
  volume={114},
  number={526},
  pages={645--656},
  year={2019},
  publisher={Taylor \& Francis}
}

@article{kim2021incremental,
  title={Incremental intervention effects in studies with dropout and many timepoints\#},
  author={Kim, Kwangho and Kennedy, Edward H and Naimi, Ashley I},
  journal={Journal of Causal Inference},
  volume={9},
  number={1},
  pages={302--344},
  year={2021},
  publisher={De Gruyter}
}

@article{schindl2024incremental,
  title={Incremental effects for continuous exposures},
  author={Schindl, Kyle and Shen, Shuying and Kennedy, Edward H},
  journal={arXiv preprint arXiv:2409.11967},
  year={2024}
}

@techreport{polley2010super,
  title={Super learner in prediction},
  author={Polley, Eric C and Van der Laan, Mark J},
  year={2010},
  month=may,
  type={U.C. Berkeley Division of Biostatistics Working Paper Series},
  number={266},
  institution={University of California, Berkeley},
  publisher={bepress}
}

@inproceedings{tsybakov2003optimal,
  title={Optimal rates of aggregation},
  author={Tsybakov, Alexandre B},
  booktitle={Learning Theory and Kernel Machines: 16th Annual Conference on Learning Theory and 7th Kernel Workshop, COLT/Kernel 2003, Washington, DC, USA, August 24-27, 2003. Proceedings},
  pages={303--313},
  year={2003},
  organization={Springer}
}

@article{liew1976inequality,
  title={Inequality constrained least-squares estimation},
  author={Liew, Chong Kiew},
  journal={Journal of the American Statistical Association},
  volume={71},
  number={355},
  pages={746--751},
  year={1976},
  publisher={Taylor \& Francis}
}

@article{schwendinger2024holistic,
  title={Holistic generalized linear models},
  author={Schwendinger, Benjamin and Schwendinger, Florian and Vana, Laura},
  journal={Journal of Statistical Software},
  volume={108},
  pages={1--49},
  year={2024}
}

@inproceedings{kim2023fair,
  title={Fair and robust estimation of heterogeneous treatment effects for policy learning},
  author={Kim, Kwangho and Zubizarreta, Jos{\'e} R},
  booktitle={International Conference on Machine Learning},
  pages={16997--17014},
  year={2023},
  organization={PMLR}
}

@book{bickel1993efficient,
  author    = {Bickel, Peter J. and Klaassen, Chris A. J. and
               Ritov, Ya'acov and Wellner, Jon A.},
  title     = {Efficient and Adaptive Estimation for Semiparametric Models},
  publisher = {Johns Hopkins University Press},
  year      = {1993}
}

@incollection{van2002semiparametric,
  title={Semiparametric statistics},
  author={van der Vaart, Aad W},
  booktitle={Lectures on probability theory and statistics (Saint-Flour, 1999)},
  pages={331--457},
  year={2002},
  publisher={Springer}
}

@article{kim2022doubly,
  title={Doubly robust counterfactual classification},
  author={Kim, Kwangho and Kennedy, Edward and Zubizarreta, Jose},
  journal={Advances in Neural Information Processing Systems},
  volume={35},
  pages={34831--34845},
  year={2022}
}

@book{tsiatis2006semiparametric,
  author    = {Tsiatis, Anastasios A.},
  title     = {Semiparametric Theory and Missing Data},
  publisher = {Springer},
  year      = {2006}
}

@article{hodson2021mean,
  title={Mean squared error, deconstructed},
  author={Hodson, Timothy O and Over, Thomas M and Foks, Sydney S},
  journal={Journal of Advances in Modeling Earth Systems},
  volume={13},
  number={12},
  pages={e2021MS002681},
  year={2021},
  publisher={Wiley Online Library}
}

@article{shapiro1990differential,
  title={On differential stability in stochastic programming},
  author={Shapiro, Alexander},
  journal={Mathematical Programming},
  volume={47},
  pages={107--116},
  year={1990},
  publisher={Springer}
}

@article{mcclean2024nonparametric,
  title={Nonparametric estimation of conditional incremental effects},
  author={McClean, Alec and Branson, Zach and Kennedy, Edward H},
  journal={Journal of Causal Inference},
  volume={12},
  number={1},
  pages={20230024},
  year={2024},
  publisher={De Gruyter}
}

@article{wang2020comprehensive,
  title={A comprehensive survey of loss functions in machine learning},
  author={Wang, Qi and Ma, Yue and Zhao, Kun and Tian, Yingjie},
  journal={Annals of Data Science},
  pages={1--26},
  year={2020},
  publisher={Springer}
}

@article{chernozhukov2014gaussian,
  title={Gaussian approximation of supremum of empirical processes},
  author={Chernozhukov, Victor and Chetverikov, Denis and Kato, Kengo},
  journal={The Annals of Statistics},
  volume={42},
  number={4},
  pages={1564--1597},
  year={2014}
}

@article{kitagawa2023treatment,
  title={Treatment choice, mean square regret and partial identification},
  author={Kitagawa, Toru and Lee, Sokbae and Qiu, Chen},
  journal={The Japanese Economic Review},
  volume={74},
  number={4},
  pages={573--602},
  year={2023},
  publisher={Springer}
}

@article{chernozhukov2016empirical,
  title   = {Empirical and Multiplier Bootstraps for Suprema of Empirical Processes of Increasing Complexity, and Related Gaussian Couplings},
  author  = {Chernozhukov, Victor and Chetverikov, Denis and Kato, Kengo},
  journal = {Stochastic Processes and their Applications},
  volume  = {126},
  number  = {12},
  pages   = {3632--3651},
  year    = {2016},
  doi     = {10.1016/j.spa.2016.04.009}
}

\end{document}